\documentclass[10pt,journal,compsoc]{IEEEtran}
\usepackage{color}
\usepackage{times}
\usepackage{epsfig}
\usepackage{graphicx}
\usepackage{algorithm}
\usepackage[noend]{algpseudocode}
\makeatletter
\def\BState{\State\hskip-\ALG@thistlm}
\makeatother
\usepackage{amsmath}
\usepackage{amssymb}
\usepackage{bbm}
\usepackage{amsxtra}
\usepackage{multirow}
\usepackage{mathtools}
\usepackage{caption}
\usepackage{cleveref}
\usepackage{url}
\usepackage{float}
\usepackage[font=small, singlelinecheck=false]{caption}
\usepackage{caption}
\usepackage[caption = false]{subfig}
\usepackage{float}
\usepackage{amsthm}
\newtheorem{theorem}{Theorem}
\newtheorem{lemma}{Lemma}

\newtheorem{corollary}{Corollary}

\newtheorem*{proof*}{Proof}
\newtheorem*{example*}{Example}
\usepackage{ragged2e}
\usepackage{xr}
\ifCLASSOPTIONcompsoc
  \usepackage[nocompress]{cite}
\else
  \usepackage{cite}
\fi
\ifCLASSINFOpdf
\else
\fi
\begin{document}
%
\title{QoE Based Revenue Maximizing Dynamic Resource Allocation and Pricing for Fog-Enabled Mission-Critical IoT Applications}
%
%
%
%

\author{Muhammad~Junaid~Farooq,~\IEEEmembership{Student Member,~IEEE,}
        and~Quanyan~Zhu,~\IEEEmembership{Member,~IEEE}
\thanks{
This research is partially supported by awards ECCS-1847056, CNS-1720230, CNS-1544782, and SES-1541164 from National Science of Foundation (NSF), and grant W911NF-19-1-0041 from Army Research Office (ARO). \newline
Muhammad Junaid Farooq is with the Department of Electrical and Computer Engineering, College of Engineering and Computer Science, University of Michigan–Dearborn, Dearborn, MI 48128 USA (E-mail: mjfarooq@umich.edu) and Quanyan Zhu is with the Department of Electrical and Computer Engineering, Tandon School of Engineering, New York University, Brooklyn, NY 11201 USA (E-mail: qz494@nyu.edu).\newline
}
}

\IEEEtitleabstractindextext{%
\begin{abstract}
\justify 
	Fog computing is becoming a vital component for Internet of things (IoT) applications, acting as its computational engine. Mission-critical IoT applications are highly sensitive to latency, which depends on the physical location of the cloud server. Fog nodes of varying response rates are available to the cloud service provider (CSP) and it is faced with a challenge of forwarding the sequentially received IoT data to one of the fog nodes for processing. Since the arrival times and nature of requests is random, it is important to optimally classify the requests in real-time and allocate available virtual machine instances (VMIs) at the fog nodes to provide a high QoE to the users and consequently generate higher revenues for the CSP. In this paper, we use a pricing policy based on the QoE of the applications as a result of the allocation and obtain an optimal dynamic allocation rule based on the statistical information of the computational requests. The developed solution is statistically optimal, dynamic, and implementable in real-time as opposed to other static matching schemes in the literature. The performance of the proposed framework has been evaluated using simulations and the results show significant improvement as compared with benchmark schemes.

\end{abstract}

\begin{IEEEkeywords}
Cloud Computing, Fog Computing, Internet of Things, Mission-Critical, Virtual Machine Instance.
\end{IEEEkeywords}}

\label{key}\maketitle

\IEEEdisplaynontitleabstractindextext

%
\IEEEpeerreviewmaketitle

\IEEEraisesectionheading{\section{Introduction}\label{sec:introduction}}

%
%
%
%

\IEEEPARstart{T}{he} interconnection of electronic sensors and actuators, known as the Internet of Things (IoT)~\cite{iot}, is creating enormous opportunities for automating systems around us and improving their efficiency. It is paving the way for the development of smart cities with active monitoring and control of public facilities, smart healthcare, smart transit systems, etc. In recent years, due to the ubiquity of the internet, there has been an increasing trend of offloading computing, control, and storage to the cloud~\cite{comput_offloading}. This is fueling the rapid growth of the IoT as it reduces the physical cost of sensors and actuators. Moreover, connectivity to the cloud opens up endless possibilities for powerful and revolutionary applications, due to the availability of massive computational power and data~\cite{promise_edge_comput}.
Therefore, cloud computing~\cite{Cloud_computing} is now becoming an integral part of the IoT ecosystem particularly for applications involving realtime analytics and Big data~\cite{big_data_iot}.

There is a wide variety of cloud enabled IoT applications that have different data processing needs. For instance, autonomous vehicles on the roads might require information about the shortest available route to the destination. On the other hand, home users might be controlling appliances remotely based on the information obtained from deployed sensors. Some of the IoT applications are highly delay-sensitive, e.g., real time systems such as those involving artificial intelligence (AI), virtual and augmented reality (VR/AR), real time control loops, streaming analytics, etc~\cite{assessment_suitability}. Such applications are referred to as \emph{mission-critical}~\cite{mc_definition} not only due to conventional `life risk' interpretation but also pertaining to the risks of public services interruption, perturbing public order, jeopardizing enterprise operation and causing losses to businesses, etc.


\begin{figure}[t!]
  \centering
  \includegraphics[width=3.6in]{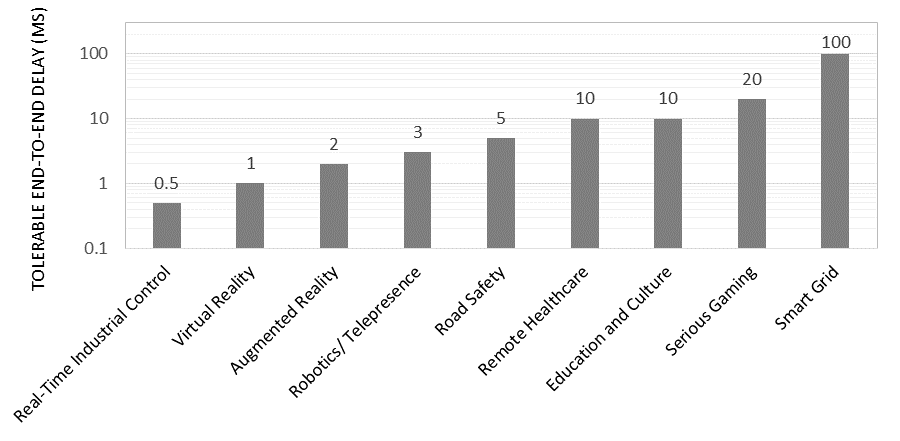}
  \caption{Tolerable end-to-end delays of some typical mission-critical IoT applications.}\label{max_latency_fig}
\end{figure}

A common characteristic of such applications is the extremely high delay sensitivity. The total delay in the response of a computational request to the cloud server, also referred to as \emph{end-to-end delay}, relies on several factors such as the latency\footnote{We use the term `latency' to refer to the delay caused due to the physical distance that is traveled by the data.}, i.e., round trip time (RTT), in addition to processing time of the computational tasks and the transmission time over the air interface. Fig.~\ref{max_latency_fig} shows the maximum tolerable end-to-end delay of some typical mission-critical IoT (MC-IoT) applications~\cite{latency_report}. It can be observed that most of the applications require an end-to-end delay of less than 10 ms. Some of them are extremely delay sensitive, requiring an end-to-end delay of 1 ms or less, such as VR and real-time industrial control applications. Although the existing cloud servers have the computational power to efficiently compute large amounts of data, the location of the server places a bottleneck on the latency. In other words, a certain time delay is inevitable regardless of the size of the task due to the distance the data has to travel before reaching back to its point of origin. Instead of sending the tremendous amounts of data, generated by the IoT to the cloud, it is more efficient if the data is analyzed at the edge of the network, i.e., close to where it is generated, to reduce the latency~\cite{minimizing_delay_iot}. Hence, a new computing architecture known as fog computing~\cite{fog_iot_cisco}, also referred to as edge computing~\cite{edge_computing_vision}, is now gaining significant attention. 


\subsection{Background \& Motivation}
\begin{figure}[t]
  \centering
  \includegraphics[width=3in]{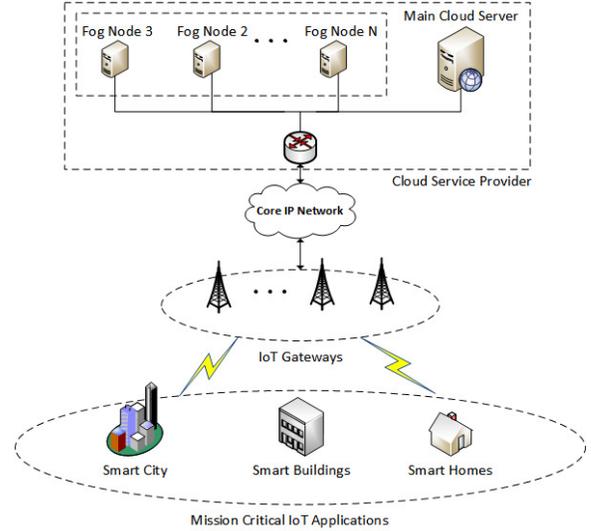}
  \caption{Architecture of a fog-enabled IoT ecosystem in which the MC-IoT devices are connected to the CSP via a gateway. The CSP has several fog nodes, equipped with a range of VMIs, in addition to the main cloud server.}\label{sys_model_fig}
\end{figure}

Fog computing is an extension of the cloud such that there are devices located at the edge of the network having computing, storage, and networking capabilities, also referred to as cloudlets or \emph{fog nodes}~\cite{fog_computing_cisco_original}. Due to the reduced distance, the availability of fog nodes can significantly reduce the response time of cloud server to incoming data~\cite{fog_and_iot}.
Hence, fog computing is emerging as one of the key enablers of fog-enabled MC-IoT~\cite{mission_critical} applications.
A cloud service provider (CSP) may have several available fog nodes in addition to the main cloud server for servicing computational requests by MC-IoT applications~\cite{hierarchical_edge_cloud_mobile}. An illustration of the hierarchical fog-cloud  architecture~\cite{hierarchical_fog_cloud} for an IoT ecosystem is provided in Fig.~\ref{sys_model_fig}. Each fog node imposes a certain delay in the response to requests coming from a certain geographical region due to its location. Furthermore, the fog nodes may have multiple virtual machine instances (VMIs), similar to the conventional cloud computing, that process the incoming data. The VMIs have different data processing capabilities according to the allocated computing resources, which results in different processing delays. Altogether, the CSP has a set of available VMIs which are characterized by their overall response time. The MC-IoT applications have a varying level of delay tolerance, i.e., the maximum delay in the response of the cloud which does not result in degradation of performance in their operation. Once an application requests the cloud for processing a computational task, the job of the CSP is to instantly make a decision of sending the received data to one of the available VMI. Since the requests by MC-IoT applications arrive at random times and have different levels of delay tolerances, the CSP needs to devise a policy according to which the VMIs are allocated and appropriately price them to maximize its revenue. 


Since the number available VMIs are limited, it is important to forward the most delay sensitive applications to the best\footnote{The notion of `best' refers to the VMI that offers the lowest end-to-end delay.} available VMI and vice versa to deliver a high quality of experience (QoE) to the users\footnote{The term `users' and `clients' are interchangeably used to refer to the client MC-IoT applications that are using the cloud for data processing.} and consequently generate higher revenue.
Note that improving the QoE of client applications enables the CSP to appropriately charge them for premium services. This implies that a highly delay sensitive applications should be allocated to a low end-to-end delay providing VMI and charged higher prices while delay tolerant applications should be allocated longer end-to-end delay providing VMIs at lower prices. However, the challenge lies in the fact that there is limited information about the nature of upcoming requests in the future.
A highly delay sensitive application may not request for service while the CSP reserves the best available VMIs, which leads to an inefficient utilization of resources resulting in lower QoE of the users. On the other hand, if the delay tolerant applications are allocated the best VMIs, then highly delay sensitive applications may request service in the future and may suffer in performance due to the unavailability of VMIs with low end-to-end delay. Secondly, as the time progresses, if the best available VMIs are not utilized while waiting for highly delay sensitive applications, the CSP may loose the opportunity of generating revenues from them at all and it may have been better to allocate them to less delay sensitive applications. Therefore, there is a need for a dynamically efficient policy that takes these trade-offs into account. We use the following example using the simplest case to elaborate the concept.





\begin{figure}[h]
	\vspace{-0.0in}
	\centering
	\includegraphics[width=3in]{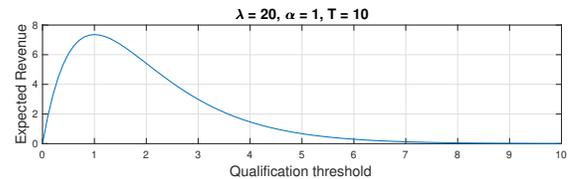}
	\caption{Expected revenue of the CSP for a single available VMI with varying qualification threshold. \vspace{-0.0in}}\label{single_revenue}
\end{figure}

\begin{example*}
Consider a CSP with only a single VMI available for allocation within a time period of 10 hours. Requests arrive sequentially at the CSP according to a Poisson process with an arrival rate of $\lambda = 20$ requests per hour. Each arriving request is assumed to have a random minimum required response rate (exponentially distributed with mean $\alpha = 1$). It implies that there is a high probability of delay tolerant applications arriving while there is a low probability of delay sensitive applications requesting service. The CSP sets a threshold specifying the barrier on the required response rate beyond which the first arriving request will be allocated to the VMI resulting in an efficiency equal to the product of the barrier and the response rate of the VMI. Consequently, a price equivalent to the qualification threshold is charged. The expected revenue of the CSP is then the product of the barrier and the probability of a qualifying request arriving.
If the barrier is set too low, the probability of a qualifying request within the allocation period will be high but the revenue generated will be low. Similarly, if the bar is set too high, then the probability of an eligible request will be low but the revenue generated if successful allocation takes place will be high. Fig.~\ref{single_revenue} shows the expected revenue against the qualification threshold set by the CSP. It is clear that there exists an optimal qualification threshold which results in maximum revenue of the CSP under uncertainty of arriving requests.
\end{example*}

Note that the example uses a static barrier to filter out computational requests. However, since the allocation period is finite, the threshold needs to be dynamic, i.e., changing with time, in order to maximize the generated revenue. Therefore, there is a need to develop an integrated policy framework that uses the QoE of the users as a basis to maximize the expected revenue generated by the CSP.


\vspace{-0.0in}
\subsection{Related Work}

\textcolor{black}{Efficient resource management in the context of cloud computing for IoT applications is currently an active area of research~\cite{resource_management_iot}. Several different objectives have been targeted for workload offloading to the fog-cloud architecture~\cite{survey_resource_management_cloud}. For instance, in~\cite{service_allocation_fog_cloud}, a framework for allocation of combined fog-cloud resources for IoT services is proposed to minimize the latency experienced by the services. On the other hand, workload allocation in fog-cloud computing for balanced delay and power consumption is provided in~\cite{workload_allocation}. Similarly, the works in~\cite{ansari_fog},~\cite{ansari_fog2}, and~\cite{ansari2} provide useful frameworks for resource provisioning in fog based systems for IoT networks. QoE has recently become an important concern in the context of cloud services~\cite{qoe_in_cloud} as opposed to the quality of service~\cite{qos_iot_fog}. It is directly linked to the latency experienced by the applications. Latency is considered as one of the biggest hurdles in traditional cloud computing for real-time and MC applications~\cite{latency_issue}. The edge-centric computing is by far the most promising approach to reducing latency~\cite{edge_centric_computing}. However, the fog nodes have limited available computational resources which need to be efficiently allocated to arriving tasks.}
%


\textcolor{black}{Resource dispatch and scheduling in cloud systems has been well studied in the literature~\cite{job_scheduling}. The research on allocation and pricing of cloud/fog resources can be broadly categorized into two main directions, i.e., static and dynamic. Both directions have been driven by several different objectives such as QoS, welfare, and revenue maximization, etc. Static allocation refers to the case where jobs are submitted and the CSP decides the allocation of resources based on the current set of jobs running on the cloud. 
Often, theoretical tools from queuing theory and optimization are leveraged to develop a scheduling and allocation policy with a specified objective.
Sometimes the decisions are made on a real-time basis, referred to as spot pricing and where the prices are set as the demand changes. However, the revenue maximization is not guaranteed over a certain time horizon. For instance, the allocation of resources for spot markets in cloud computing environments has been explored in~\cite{spot_markets}.
Most such works consider a matching problem between requests and virtual machines, which is not a sequential problem as complete information is available at the time of decision.}

\textcolor{black}{The idea of dynamic pricing and revenue maximization in the presence of stochastic demand has also been investigated in the context of cloud computing~\cite{cloud_pricing}. However, it does not use price discrimination for differentiated services offered by the cloud which is imperative in the case of fog-enabled MC-IoT.  Revenue maximization if often accomplished by the use of auctions to extract the maximum revenue under hidden type of the applications~~\cite{mechanism_design_cloud}. However, auctions are not instantaneous, even though they may be implemented in real-time~\cite{static_auction_fogspot}, \cite{bid_the_cloud} \cite{auction_based}. Revenue maximization in a dynamic setting is sparingly investigated in relevant literature due to its increased complexity~\cite{online_job_dispatching}. However, they do not focus on the MC-IoT applications and the revenue maximization aspect. Those that consider real-time decision making, do not consider a finite time horizon to adapt the decision according to the remaining allocation time. This is why we resort to statistical analysis 
to dealing with the cloud resource provisioning and pricing decisions for real-time and mission-critical needs of the applications. The pricing and allocation is based on the QoE provided to the users, which in turn is based on the delay tolerance of the applications as well as the response rates of the VMIs. In this work, we consider the case where the dynamics of the allocation process are taken into account. In other words, the sequential arrival, instantaneous allocation, incomplete information, and allocation deadline are the salient features that distinguish this work from the literature.
}







In the following subsection, we highlight the key contributions of this paper in comparison with the existing literature.

\subsection{Contributions}

Most existing work on job dispatching and scheduling is focused on the static assignment of tasks to computational resources in the fog-cloud environment. For instance, a task offloading framework using matching theory has been proposed in~\cite{matching_theory}. However, it is based on static pairing between tasks and fog nodes and the dynamic aspects of task arrival are not considered. Similarly, an index based task assignment and scheduling of tasks is proposed in~\cite{index_task_assignment}. Several works do consider real-time task allocation and dispatch. For instance, an online job dispatching and scheduling algorithm is proposed in~\cite{online_job_dispatching}, where jobs are released in arbitrary order and times by mobile devices and offloaded to unrelated servers with both upload and download delays. However, the approach is based on a greedy algorithm and is not strategic, i.e., does not make use of the statistical information about computational requests to make more effective allocation decisions. Furthermore, pricing and revenue maximization has been looked into sparingly.

In this paper, we present a revenue maximizing perspective towards allocation and pricing in fog based systems designed for mission critical IoT applications. The QoE resulting from the pairing of fog resources with computation requests is used as a basis for pricing. We develop a dynamic policy framework leveraging the literature in economics, mechanism design~\cite{revenue_maximization}, and dynamic revenue maximization~\cite{revenue_book} to provide an implementable mechanism for dynamic allocation and pricing of sequentially arriving IoT requests that maximizes the expected revenue of the CSP.
The developed optimal policy framework assists in both determining which fog node to allocate an incoming task to and the price that should be charged for it for revenue maximization. The proposed policy is statistically optimal, dynamic, i.e., adapts with time, i.e., and is implementable in real-time as opposed to other static matching schemes in the literature. The dynamically optimal solution can be computed offline and implemented in real-time for sequentially arriving computation requests.

The rest of the paper is organized as follows: Section~\ref{Sec:Model} provides a description of the system model, Section~\ref{Sec:Main_Results} provides the details of the dynamically optimal mechanism for QoE based revenue maximization framework along with the main results, and Section~\ref{Sec:Numerical_Experiments} presents the results of numerical experiments with discussions. Finally, Section~\ref{Sec:Conclusion} concludes the paper.

\section{Model Description}\label{Sec:Model}

We consider a CSP in a fog enabled IoT ecosystem having a set of $k$ available fog nodes in addition to the main cloud server and serving a certain geographical region containing MC-IoT devices. The fog nodes have an associated average latency denoted by $l_i \in \mathbb{R}, i \in \{1, \ldots, k\}$ which depends on the distances of the fog nodes from the locations of origin of the processing requests. The number of available VMIs at the fog nodes is denoted by $n_i, i \in \{1, \ldots, k\}$ and each of the available VMI is characterized by its processing delay for a fixed number of computational operations denoted by $\tau_{ij}^{(p)}, i = \{1, \ldots, k\}, j \in \{1, \ldots, n_i\}$. There are a total of $\sum_{i=1}^{k}n_i = N$ VMIs available for allocation by the CSP to sequentially arriving MC-IoT requests. The latency of each individual VMI is denoted by $\tau^{(l)}_{ij} = l_i, \forall j = \{1, \ldots, n_i\}$. The end-to-end delay offered by the VMIs can be evaluated as $\tau_{ij}^{(l)} + \tau_{ij}^{(p)} + \tau^{(o)}$, where $\tau^{(o)}$ represents the other delays including the transmission delay over the air interfaces. Consequently, the average response rate of the $j^{\text{th}}$ VMI at compute node $i$ can be expressed as follows\footnote{Performance metrics and utility of similar forms are used in the literature for fog-enabled IoT systems, e.g.,~\cite{matching_theory,iot_service_delay}.}:
\begin{align}
r_{ij} = \frac{1}{\tau_{ij}^{(l)} + \tau_{ij}^{(p)} + \tau^{(o)}}.
\end{align}
With an abuse of notation, the average response rates of the available VMIs can further be denoted by $\mathbf{r} = [r_1, r_2, \ldots, r_N]$, where $r_1 \geq r_2 \geq \ldots \geq r_N$, and $\mathcal{M}$ defines the mapping of values of the pair $(i,j), i \in \{1, \ldots, k\}, j \in \{1, \ldots, n_i\}$ from the original $r_{ij}$ to the set $n = \{1, \ldots, N\}$ in the new response rates denoted by $r_n$. It implies that the VMI corresponding to a response rate $r_1$ is the best in terms of end-to-end delay for processing IoT data while the one corresponding to $r_N$ is the least favorable in terms of end-to-end delay.

MC-IoT devices in a given geographical region are connected to the CSP via a set of IoT gateways. We assume that requests for processing by MC-IoT applications arrive sequentially at the CSP according to a homogeneous Poisson process with intensity $\lambda$. Each request is characterized by its maximum tolerable delay for successful operation denoted by $d_i, i \geq 1$. In other words, there is a minimum required response rate by the applications. Upon arrival of a request by an MC-IoT application, the CSP has to make a decision to forward the data to one of available VMIs at the fog nodes. An illustration of the system model is provided in Fig.~\ref{sys_model_fig}.

It is pertinent to mention that before the actual task offloading, the task metadata is sent to the cloud server for the allocation decision. Furthermore, we assume that the allocation decision is instantaneous and the delay in transmitting metadata is not relevant in the computation of delay experienced in the transmission and processing of actual application data since it occurs only once before the allocation is made. 
The allocation is made for long term access of the resources by the applications. The applications can release the resources anytime based on their requirements. Once the resources are returned, they can be reallocated to other incoming applications. The proposed allocation framework allows this since the solution is dynamically optimal. In other words, it depends on the state, which in this case is the number of VMIs available. Therefore, if the optimal policies have been precomputed for multiple available VMs, then the policy can be switched in real-time and it would still be optimal for the remaining time available.

\subsection{Allocation Efficiency and Quality of Experience}
Upon arrival of a request by an MC-IoT application, the CSP determines the sensitivity of the application to a delay in the response and is subsequently required to allocates it to one of the available VMIs at the fog nodes while charging a particular price. The VMIs are allocated to requesting applications for long term and therefore, once a VMI is allocated to an application, it becomes unavailable for allocation to other applications in the future\footnote{Note that this assumption is only made for obtaining an optimal dynamic policy using an open loop methodology. In practical implementation, the policy can be adapted based on how many VMIs are available.}. Since the VMIs are perishable, i.e., they have no value if not successfully allocated within the allocation time horizon $T$, so it is important for the CSP to optimally allocate the available resources within a certain time frame\footnote{The time horizon refers to the period over which the allocation has to occur which can be related to the demand window.}. It also motivates the idea of dynamic pricing, i.e., to charge higher prices earlier and reduce them as time goes on for maximizing the revenue of the CSP.

Let $x_i = d_i^{-1}$ be the minimum required response rate by the $i^{\text{th}}$ application.
Each application reports this characteristic to the CSP at the time of requesting service. We assume that each $x_i \in \mathbb{R}, i \geq 1,$ is an independent and identically distributed (i.i.d.) random variable with a probability density function (pdf) denoted by $f_X(x)$ and a cumulative distribution function (cdf) denoted by $F_X(x)$\footnote{We assume that the probability distribution of the arriving tasks is known a priori. This is done for analytical tractability and policy development as data-driven approaches are prohibitive in terms of obtaining an implementable dynamic policy.}.
The product $x_ir_j$ can be used as a measure of efficiency when the $i^{\text{th}}$ application is allocated to the $j^{\text{th}}$ available VMI. The QoE as a result of the pairing, denoted by $\Phi(x_i,r_j)$ is quantified as follows:
\begin{align}
\Phi(x_i,r_j) = (x_i r_j)^{\frac{1}{\eta}}, \ \ i,j \in \{1, \ldots, N\},
\end{align}
where $\eta \geq 1$ is a constant controlling the rate of increase in QoE with respect to the efficiency of allocation. The concave nature of the QoE function implies that an increase in allocation efficiency results in diminishing improvements in the QoE of the applications. \textcolor{black}{Note that the choice of utility function is not exclusive and any arbitrary function with similar structural properties, i.e., concave and increasing in both the average response rate and the minimum required rate of the applications , can be used. However, the choice here is made for simplicity and tractability of analysis inspired by the Cobb-Douglas function from econometrics~\cite{cobb_douglas} that models the production output relationship between the amounts of two or more inputs.}

In this section, we provide a framework for dynamically maximizing the expected revenue of the CSP based on the QoE of the users. We focus on a mechanism design approach, which is based on developing optimal implementable policies for allocation and pricing. A direct mechanism is provided whereby each requesting application reports its minimum required response rate and the CSP allocates one of the available VMIs to it. For the optimal allocation to be implementable, an allocation policy and a payment rule is required that is incentive compatible\footnote{Incentive compatibility is a concept from mechanism design theory that ensures that no agent has an incentive to misreport its privately known characteristic.}~\cite{Myerson} in the presence of individually rational users. We first state an allocation rule which satisfies the aforementioned conditions and then provide a pricing strategy that subsequently implements the allocation.
\vspace{-0.0in}
\subsection{Allocation Policy}\label{Sec:Allocation}

Based on the required response rate of each randomly arriving computational request by the IoT devices, the CSP has to allocate a VMI at one of the fog nodes to maximize the expected QoE of the users. It is clear that an application that requires a high response rate should be allocated to the VMI offering a high response rate and vice versa for efficient utilization of resources. However, the problem rests in the fact that an allocation decision has to be made without knowledge of the type of applications that will request in the future. It is shown in the literature~\cite{albright} that a dynamically efficient allocation policy in such situations can be achieved using a partition on the characteristic of the sequentially arriving agent. We provide a deterministic and Markovian allocation policy $\pi_t(x, \mathbf{r}_t)\footnote{Throughout the paper, the subscript $t$ is used to refer to the time dependence.}: \mathbb{R} \rightarrow \mathbf{y}_t$, i.e., at each time $t$, a fixed non-random policy is used that only depends on the current time instant and set of available VMIs at time $t$. 
The key result in the allocation policy is provided by the following theorem.
\begin{lemma}[Adapted from~\cite{albright}] \label{theorem_allocation}
	A deterministic and Markovian policy at time $t$, i.e., $\pi_t(x, \mathbf{r}_t)$ is implementable iff there exists a set of functions $y_{i}(t)$, $i = 1, \ldots, N_t$, such that $0 < y_{N_t}(t) < y_{N_t-1}(t) < \ldots < y_{1}(t) < y_{0}(t) = \infty$. The allocation policy is such that $\pi_t(x, \mathbf{r}_t) = r_i$ if $x \in [y_{i}(t), y_{i-1}(t)]$ and $\pi_t(x, \mathbf{r}_t) = \emptyset$ if $x < y_{N_t}(t)$.
\end{lemma}
The allocation policy depends on the nature of the requesting applications as well as the available VMIs at time $t$. Lemma~\ref{theorem_allocation} implies that if at time $t$, the requesting application has a required response rate lower than $y_{N_t}(t)$, then the CSP will not allocate any compute node to the application as it aspires to save the VMIs for higher valued application requests in the future. However, if the required response rate is between $y_{i}(t)$ and $y_{i - 1}(t)$, then the $i^{th}$ best available VMI is allocated to the requesting application. It can be observed that the number of partitions or cutoff curves depends on the set of available compute nodes at time $t$.
In the subsequent subsections, we provide a suitable pricing scheme associated with the aforementioned policy and provide the optimal time varying cutoff values for efficient allocation.



\subsection{Pricing Policy}
For the proposed allocation policy, there is a need to appropriately price the applications for their achieved QoE. We assume that the MC-IoT applications are individually rational, i.e., no application will be willing pay more than the QoE it achieves by using the allocated VMI. The partition based allocation policy described in Section~\ref{Sec:Allocation} provides a natural mechanism for pricing the applications for their proposed allocations. Since an application with a required response rate $x \in [y_i(t), y_{i-1}(t)]$ at time $t$ is allocated to a VMI with a response rate $r_i$, it has achieved an improvement in QoE of atleast $(r_i^{\frac{1}{\eta}} - r_{i+1}^{\frac{1}{\eta}})y_i^{\frac{1}{\eta}}(t)$ as compared to the next best allocation. Therefore, it must pay an equivalent price to be allocated to a VMI with response rate $r_i$ as compared to the one with $r_{i+1}$. This process can be continued recursively to obtain implementable prices for each of the available VMIs. It has been shown in literature~\cite{albright} that such incentive compatible pricing structure in which the price is selected on the basis of displaced value is implementable. Therefore, the optimal prices can be completely determined by the implementation conditions. The price charged to an MC-IoT application that is allocated to a $j^{\text{th}}$ best VMI, i.e., $x \in \left[ y_{j}(t), y_{j-1}(t) \right)$, at time $t$ can be expressed as follows:
\begin{align}
P_j(\mathbf{r}_t,t) = \sum_{i = j}^{N_t} \left( r_i^{\frac{1}{\eta}} - r_{i+1}^{\frac{1}{\eta}} \right) y_{i}^{\frac{1}{\eta}}(t),
\end{align}
Note that $r_i > r_{i+1}, \forall i = 1, \ldots, N-1$ due to the initial ordering. The pricing policy is progressive in a relative sense with the lowest QoE achieved as a reference.
We provide a simple example to further elaborate the pricing policy. If there is a single VMI available, the price would be $P_1(\{r_1\},t) = r_1^{\frac{1}{\eta}} y_1^{\frac{1}{\eta}}(t)$, which is equivalent to the QoE of the application to which the VMI was allocated. However, if there are two VMIs avilable, the prices for the lower response rate VMI is set to be $P_2 ( \{ r_1,r_2\}, t ) = r_2^{\frac{1}{\eta}} y_2^{\frac{1}{\eta}}(t)$ and the price for the higher response rate VMI is set to be $P_1(\{ r_1,r_2\}, t ) = r_2^{\frac{1}{\eta}} y_2^{\frac{1}{\eta}}(t) +  (r_1^{\frac{1}{\eta}} -  r_2^{\frac{1}{\eta}}) y_1^{\frac{1}{\eta}}(t)$ Note that the price for the latter is simply the price of the former plus the improvement in QoE experienced by getting a VMI with response rate $r_1$ instead of $r_2$. In other words, the lowest response rate VMI is priced at the base price equivalent to the QoE achieved by its allocation. The next higher one is priced at the base price plus the improvement in QoE achieved as a result of being allocated a better available VMI.




\section{Main Results}\label{Sec:Main_Results}









In this section, the goal is to maximize the expected revenue generated by the CSP using the pricing strategy developed in Section~\ref{Sec:Model}. We first begin with solving the problem for the case of a single available VMI and then generalize it to multiple VMIs using a recursive approach.

\subsection{Single VMI Case}

If only a single VMI is available to the CSP with a response rate $r_1$, then the expected revenue generated by its allocation within a time period $T$ can be expressed as follows:
\begin{align}
R(\{r_1\},t) &=  \int_t^T P_1(\{r_1\},t) h_{1}(s) ds =  r_1^{\frac{1}{\eta}} \int_t^T y_{1}^{\frac{1}{\eta}}(s) h_{1}(s) ds, \label{revenue_single_expr}
\end{align}
where $h_1(s)$ is the probability density of waiting time until the first arrival of a request with a required response rate of greater than $y_1(s)$.
The objective of the CSP is to determine the optimal allocation threshold $y_1^{*}(t)$ such that the expected revenue is maximized, i.e.,
\begin{align}
y_1^{*}(t) = \arg \max  r_1^{\frac{1}{\eta}} \int_t^T y_{1}^{\frac{1}{\eta}}(s) h_{1}(s) ds.
\end{align}
Note that the revenue expression in~\eqref{revenue_single_expr} involves an integral of a quasi-concave function since for a fixed time $t \geq 0$, the function $y_1^{\frac{1}{\eta}}(t)$ is monotonically increasing while $h_1(t)$ is monotonically decreasing in $y_1(t)$. Hence, a unique maxima can be obtained for every $t$. However, finding the optimal function requires a variational approach.
The optimal dynamic threshold for the case of a single VMI can be determined using the following theorem.
\begin{theorem} \label{single_theorem}
	If the CSP has a single available VMI at time $t$, then it is optimal to allocate it to a requesting application if the required delay tolerance $x_i \geq y_1^*(t)$, where the optimal threshold $y_1^*(t)$ solves the following equation:
	\begin{align}
	y_1^*(t) =& \left( \frac{1 - F_{\hat{X}}((y_1^*)^{\frac{1}{\eta}}(t))}{f_{\hat{X}}((y_1^*)^{\frac{1}{\eta}}(t))} + \right. \notag \\ &\left.  \lambda \int_t^T \frac{(1 - F_{\hat{X}}((y_1^*)^{\frac{1}{\eta}}(s)))^2}{f_{\hat{X}}((y_1^*)^{\frac{1}{\eta}}(s))} ds \right)^{\eta}
	\end{align}
	\begin{proof}
		See \textbf{Appendix~\ref{proof_single_theorem}}.
	\end{proof}
\end{theorem}

In the subsequent subsection, we extend the approach to the case of multiple available VMIs.

\subsection{Multiple VMI Case} \label{Sec:cutoff}

In this subsection, we first present the case of two available VMIs at the CSP and then generalize it to the case of multiple available VMIs. If there are only two available VMIs with response rates $r_1$ and $r_2$, then the expected revenue of the CSP can be expressed as follows:
\begin{align}
R(\{r_1,r_2\},t) &= \int_0^T \left( P_2(\{r_1,r_2\}t) + R(\{r_1\},t) \right)  h_2(t) dt + \notag \\  &\int_0^T \left( P_1(\{r_1,r_2\}t) + R(\{r_2\},t) \right) h_1(t) dt,\notag \\
& = \int_0^T \left(  r_2^{\frac{1}{\eta}} y_2^{\frac{1}{\eta}} + R(r_1,t)\right) \times \notag \\ &\lambda (1 - F_{\hat{X}}(y_2^{\frac{1}{\eta}}(t))) e^{-\int_0^t \lambda (1 - F_{\hat{X}}(y_2^{\frac{1}{\eta}}(s)))ds } dt + \notag \\ &(r_1^{\frac{1}{\eta}} - r_2^{\frac{1}{\eta}}) \int_0^T \left( y_1^{\frac{1}{\eta}}(t) - R(1,t) \right) \times \notag \\
&\lambda (1 - F_{\hat{X}}(y_1^{\frac{1}{\eta}}(t))) e^{ - \int_0^t \lambda (1 - F_{\hat{X}}(y_2^{\frac{1}{\eta}}(s)))ds} dt.
\end{align}
where $h_1(t)$ represents the density of waiting time till the first arrival of a request with a required response rate of atleast $y_1(t)$ if no request with a required response rate in the interval $[y_2(t), y_1(t))$ has arrived. Similarly, $h_2(t)$ represents the density of waiting time till the first arrival of a request with a required response rate in the interval $[y_2(t), y_1(t))$ if no request with a required response rate in the interval $[y_1(t), \infty )$ has arrived. The optimization problem in this case becomes the following.
\begin{align}
y_2^*(t) &= \arg \max \int_0^T \left(  r_2^{\frac{1}{\eta}} (y_2^*)^{\frac{1}{\eta}} + R(r_1,t)\right) \times \notag \\ &\lambda (1 - F_{\hat{X}}((y_2^*)^{\frac{1}{\eta}}(t))) e^{-\int_0^t \lambda (1 - F_{\hat{X}}((y_2^*)^{\frac{1}{\eta}}(s)))ds } dt + \notag \\ &(r_1^{\frac{1}{\eta}} - r_2^{\frac{1}{\eta}}) \int_0^T \left( (y_1^*)^{\frac{1}{\eta}}(t) - R(1,t) \right) \times \notag \\
&\lambda (1 - F_{\hat{X}}((y_1^*)^{\frac{1}{\eta}}(t))) e^{ - \int_0^t \lambda (1 - F_{\hat{X}}((y_2^*)^{\frac{1}{\eta}}(s)))ds} dt.
\end{align}
The optimal dynamic threshold for the case of two available VMIs can be determined using the following theorem.
\begin{theorem}
	If the CSP has a two available VMIs at time $t$, then it is optimal to allocate the low response rate VMI to a requesting application if the required delay tolerance $x_i \in [y_2^*(t), y_1^*(t))$, and to allocate the high response rate VMI if the required delay tolerance $x_i \geq  y_2^*(t)$  where the optimal threshold $y_2^*(t)$ solves the following equation:
	\begin{align}
	y_2^*(t) = \Bigg( \frac{1 - F_{\hat{X}}((y_2^*)^{\frac{1}{\eta}}(t))}{F_{\hat{X}}((y_2^*)^{\frac{1}{\eta}}(t))} + &\lambda \int_t^T \frac{(1 - F_{\hat{X}}((y_2^*)^{\frac{1}{\eta}}(s)))^2}{F_{\hat{X}}((y_2^*)^{\frac{1}{\eta}}(s))} ds   \notag \\  & - R(1,t) \Bigg)^{\eta}.
	\end{align}
	\begin{proof}
		See~\textbf{Appendix~\ref{proof_allocation_theorem}}.
	\end{proof}
\end{theorem}
Note that the optimal threshold $y_2^*(t)$ relies on obtaining the threshold $y_1^*(t)$. In the general case, it can be shown by induction that the optimal thresholds can be obtained recursively using the following theorem:
\begin{theorem} \label{general_theorem}
	If there are $N_t$ available VMIs at time $t$, then it is optimal to allocated a VMI with response rate $r_i$ to an incoming request with minimum required response rate $x_i \in [y_i^*(t), y_{i-1}^*(t)]$, where the optimal dynamic thresholds $y_i^*(t)$
	satisfy the following recursive equation:
	\begin{align}\label{eq_y_general}
	y_i^*(t) =   & \Bigg( \frac{1 - F_{\hat{X}}(((y_i^*)^{\frac{1}{\eta}})(t))}{f_X(((y_i^*)^{\frac{1}{\eta}})(t))} + \notag \\ & \lambda \int_t^T  \frac{(1 - F_{\hat{X}}(((y_{i-1}^*)^{\frac{1}{\eta}})(s)))^2}{f_X(((y_{i-1}^*)^{\frac{1}{\eta}})(s))} ds \ \ - \notag \\ &  \lambda  \int_t^T \frac{(1 - F_{\hat{X}}(((y_{i}^*)^{\frac{1}{\eta}})(s)))^2}{f_X(((y_{i}^*)^{\frac{1}{\eta}})(s))} ds \Bigg)^{\eta},  i = 2, \ldots, N_t.
	\end{align}
	\begin{proof}
		See~\textbf{Appendix~\ref{proof_general_theorem}}.
	\end{proof}
\end{theorem}
Note that the optimal cutoff curves are independent of the response rates of the available VMIs. In fact, they depend only on the number of available VMIs at time $t$ and on the statistical information about the sequentially arriving computational requests by MC-IoT applications. In the following set of corollaries, we provide the results obtained for special cases of the statistical information about arriving requests.






\begin{corollary}\label{corol_exp}
\label{exponential_derivation}
Assuming that the transformed required response rate of sequentially arriving MC-IoT applications, denoted by $\hat{X}$ (See Appendix~\ref{single_theorem}, follows an exponential distribution with a mean of $\alpha^{-1}$, i.e., $f_{\hat{X}}(x) = \alpha e^{-\alpha x}$, and $F_{\hat{X}}(x) = 1 -  e^{- \alpha x}$, then the optimal cutoff curves for allocation can be expressed as follows:
\begin{align}
y_1^*(t) = & \frac{1}{\alpha^{\eta}} \left[1 +  \log \left( 1 + \frac{\lambda (T - t)}{e}  \right)   \right]^{\eta},\\
y_2^*(t) = & \frac{1}{\alpha^{\eta}} \left[1 + \log \left( 1 + \frac{\lambda^2 (T - t)^2}{2e \left(\lambda (T - t) + \mathrm{e} \right)} \right) \right]^{\eta}, \\
y_3^*(t) = &\frac{1}{\alpha^{\eta}} \hspace{-0.05cm} \left[1\hspace{-0.05cm} + \hspace{-0.05cm} \log \hspace{-0.1cm}\left( \hspace{-0.05cm}1 \hspace{-0.05cm}+ \hspace{-0.05cm}\frac{\lambda^3 (T - t)^3 }{3e \left( \lambda^2 (T - t)^2 + 2 \mathrm{e}( \lambda (T - t) + e )  \right)}\hspace{-0.1cm} \right) \hspace{-0.05cm} \right]^{\eta}\hspace{-0.1cm},
\end{align}
The remaining lower cutoff curves cannot be easily derived in closed form and can be computed numerically from (10) using the Picard iterative process~\cite{Picard}.
\end{corollary}

\begin{corollary}
Assuming that the transformed required response rate of sequentially arriving MC-IoT applications, denoted by $\hat{X}$ (See Appendix~\ref{single_theorem}), is uniformly distributed in the interval $[0, \beta]$, i.e., $f_{\hat{X}}(x) = \frac{1}{\beta}$ and $F_{\hat{X}}(x) = \frac{x - \beta }{\beta}, x \in [0, \beta]$, then the optimal cutoff curves for the allocation of the best available VMI can be expressed as follows: \vspace{-0.0in}
\begin{align}
y_1^*(t) = \beta^{\eta}\left(1  -  \frac{2}{\lambda (T - t) + 4} \right)^{\eta}.
\end{align}
The lower cutoff curves cannot be easily obtained analytically and thus require numerical methods such as the Picard iterative process~\cite{Picard}.
\end{corollary}

\subsection{Expected Revenue}



\begin{figure*}[t]
  \centering
  \includegraphics[width=4.5in]{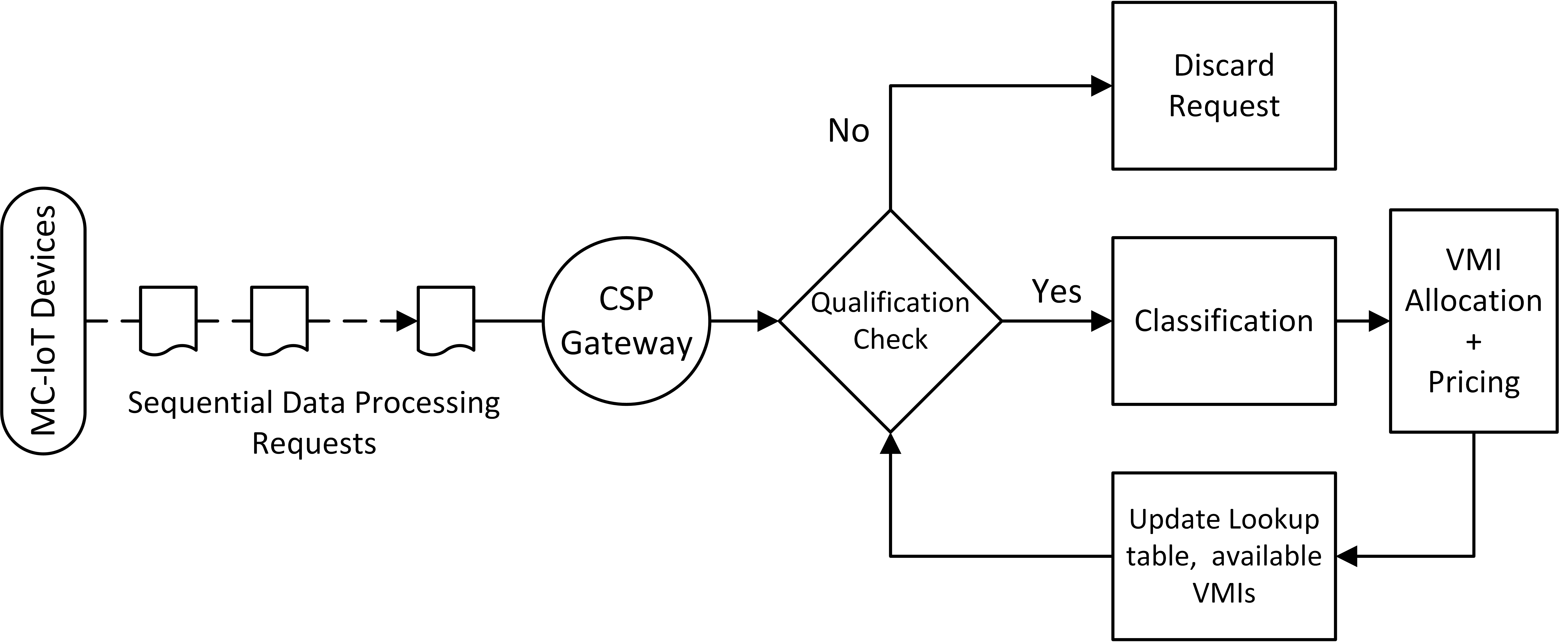}
  \caption{Resource Allocation and Pricing Flow Diagram.}\label{flow_fig}
\end{figure*}


The expected revenue from the optimal allocation of VMIs to sequentially arriving MC-IoT requests can be expressed by the following theorem:
\begin{theorem}
The expected revenue of the CSP at time $t$ if a total of $N_t$ VMIs with response rate defined by $\mathbf{r}_t = [r_1, r_2, \ldots, r_{N_t}]$ can be expressed as:
\begin{align}
\mathcal{R}(\mathbf{r}_t,t) = \sum_{i=1}^{N_t} r_i^{\frac{1}{\eta}} \left(  y_i^{\frac{1}{\eta}}(t) - \frac{1 - F_{\hat{X}}(y_i^{\frac{1}{\eta}}(t))}{f_X(y_i^{\frac{1}{\eta}}(t))} \right).
\end{align}
\begin{proof}
The expected revenue if only a single VMI with unit response rate is available is given by $\mathcal{R}(\{1\},t) =  \int_t^s \frac{(1 - F_{\hat{X}}(y_1^{\frac{1}{\eta}}(s)))^2}{f_X(y_1^{\frac{1}{\eta}}(s))}ds =  \left( y_1^{\frac{1}{\eta}}(t) - \frac{1 - F_{\hat{X}}(y_1^{\frac{1}{\eta}}(t))}{f_X(y_1^{\frac{1}{\eta}}(t))} \right).$ Similarly, if two VMIs with unit response rates are available, then using Theorem~\ref{single_theorem}, the expected revenue is given by $\mathcal{R}(\{1,1\},t) =  \int_t^s \frac{(1 - F_{\hat{X}}(y_2^{\frac{1}{\eta}}(s)))^2}{f_X(y_2^{\frac{1}{\eta}}(s))}ds =  \left( y_1^{\frac{1}{\eta}}(t) - \frac{1 - F_{\hat{X}}(y_1^{\frac{1}{\eta}}(t))}{f_X(y_1^{\frac{1}{\eta}}(t))}\right) + \left( y_2^{\frac{1}{\eta}}(t) - \frac{1 - F_{\hat{X}}(y_2^{\frac{1}{\eta}}(t))}{f_X(y_2^{\frac{1}{\eta}}(t))}\right).$ Using an inductive argument along with the fact that $\mathcal{R}(\{r_j\},t) = r_j^{\frac{1}{\eta}} \mathcal{R}(\{1\},t)$ proves the result.
\end{proof}
\end{theorem}
Notice that the expected revenue is linear in the response rates of the available VMIs and increases if a high response rate or equivalently a low latency is provided by the VMIs at the fog nodes.


\subsection{Implementation of Dynamic VMI Allocation and Pricing}
In this section, we explain the operation of the proposed QoE based revenue maximizing dynamic allocation and pricing framework. Requests for remote computations by MC-IoT applications arrive at the CSP at random times and have a delay sensitivity which is unknown \emph{a priori}. Based on the initial number of available VMIs at the CSP $N_0$, there is a minimum cutoff threshold $y_{N_0}(t)$ which an incoming request has to cross before being allocated a VMI. If the incoming request qualifies for an allocation, the CSP needs to decide which VM to allocate to the user. A lookup table denoted by $\mathcal{T}$ is prepared by the CSP which contains pre-computed optimal dynamic cutoff curves determined in Section~\ref{Sec:cutoff}. Using this lookup table, the request is optimally classified for allocation to one of the available VMIs. The data from the MC-IoT request is forwarded to the selected VMI and a price, also available in the lookup table, is charged to the requesting application. Once the allocation has been completed, the VMI is removed from the set of available VMIs. The lookup tables are updated by removing the least cutoff threshold $y_{N_0}(t)$. The remaining available VMIs are re-arranged in descending order and their prices are updated in the lookup table $\mathcal{T}$. This process is repeated until either all the available VMIs have been successfully allocated or the allocation period has ended.
Note that the lookup tables are useful in the event that an application decides to leave the cloud. Once an application has released the allocated VMI, it becomes available again for re-allocation. Hence, the allocation policy can simply be updated by moving on to the threshold with higher number of available VMIs. This is because the solution is dynamically optimal and can be updated on the go if the state, i.e., the number of available VMIs changes. This procedure has been summarized in \textbf{Algorithm~\ref{alg1}} and the associated flow diagram is provided in Fig.~\ref{flow_fig}.

\begin{algorithm}
\caption{Dynamic VMI Allocation and Pricing}\label{alg1}
\begin{algorithmic}[1]
\Require{Initialize counter = 0, request index $i=1$, starting time $t = 0$, $\mathbf{r}_0 = \{r_i, i = 1, \ldots, N_0: r_1 \geq r_2 \geq \ldots \geq r_{N_0}\}$.}
\While{counter $< N_0$ \textbf{ and } $t < T$}
\State {\begin{tabular}{@{\hspace*{0.5em}}l@{}}Determine the required response rate\\ of the arriving computational request $x_i$.\end{tabular}}
\If{ $x_i \geq y_{N_0 - \text{counter}}(t)$}
\State {\begin{tabular}{@{\hspace*{0.5em}}l@{}}Classify the request using the lookup table, i.e.,\\ determine $j : x_i \in [y_j(t), y_{j-1}(t)]$.\end{tabular}}
\State {\begin{tabular}{@{\hspace*{0.5em}}l@{}}The $i^{\text{th}}$ requesting application is allocated to the\\ $j^{\text{th}}$ highest VMI.\end{tabular}}
\State {\begin{tabular}{@{\hspace*{0.5em}}l@{}}Use the mapping $\mathcal{M}$ to realize the allocation\\ in terms of the fog node and available VMI.\end{tabular}}
\State {\begin{tabular}{@{\hspace*{0.5em}}l@{}}Use $\mathcal{T}$ to charge a price $P_j(\mathbf{r}_t)$ to the requesting\\ application.\end{tabular}}
\State {\begin{tabular}{@{\hspace*{0.5em}}l@{}}Remove the VMI corresponding to $r_j$ from the list\\ of  available VMIs.\end{tabular}}
\State {\begin{tabular}{@{\hspace*{0.5em}}l@{}}Re-arrange the set of available VMIs in \\ descending order.\end{tabular}}
\State {\begin{tabular}{@{\hspace*{0.5em}}l@{}}Update lookup table for relevant cutoff curves\\ $\mathcal{T} = \mathcal{T} \backslash y_{N_0 - \text{counter}}(t)$.\end{tabular}}
\State {\begin{tabular}{@{\hspace*{0.5em}}l@{}}Update $\mathcal{T}$ with new prices corresponding to the\\ updated list of available VMIs.
      \end{tabular}}
\EndIf
$i \gets i+1$.
\EndWhile
\end{algorithmic}
\end{algorithm}


\begin{figure*}[t!]
\centering
\vspace{-0.0in}
\subfloat[]{\includegraphics[width = 3.1in]{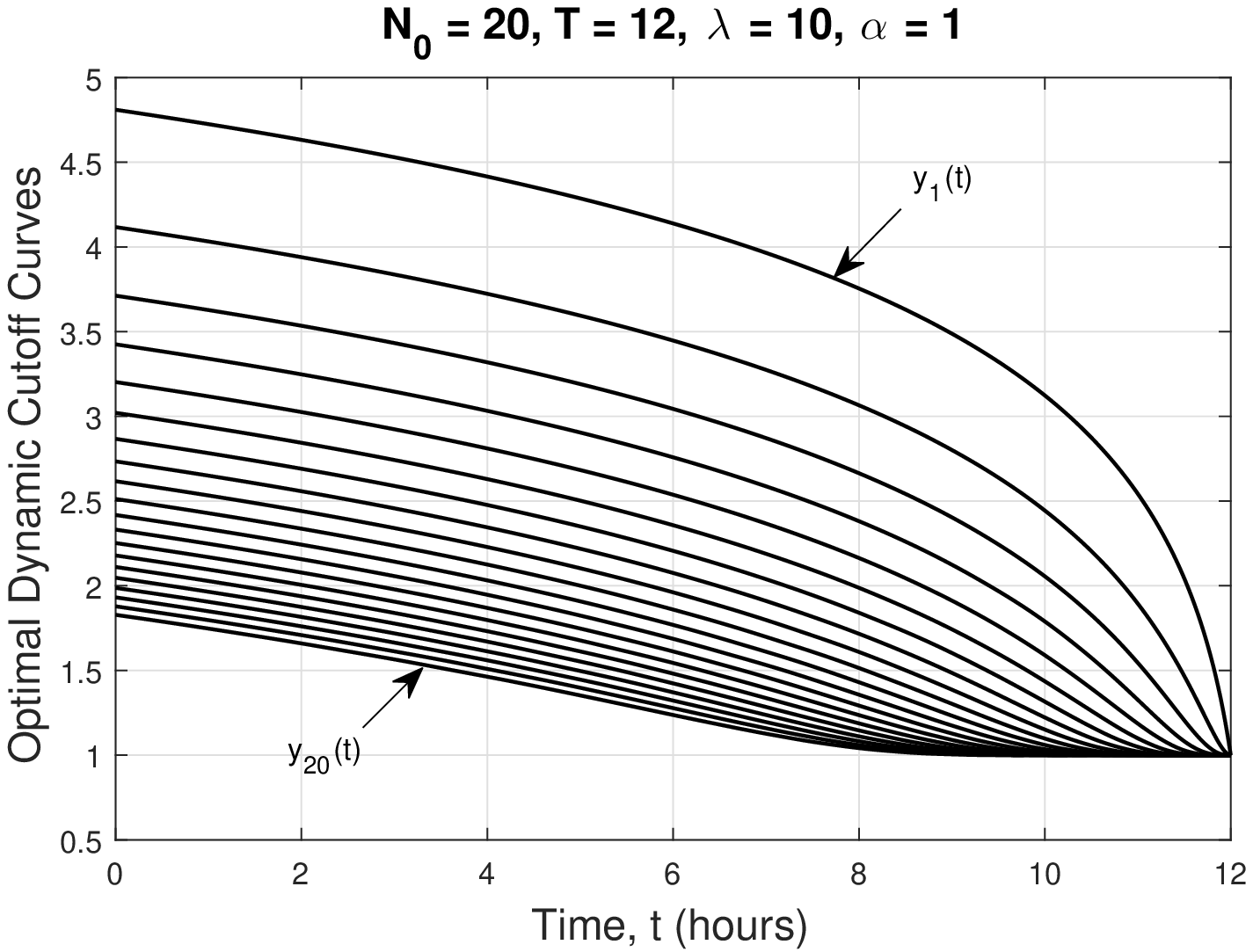} \label{cutoff_exp_fig}} \ \
\subfloat[]{\includegraphics[width = 3.1in]{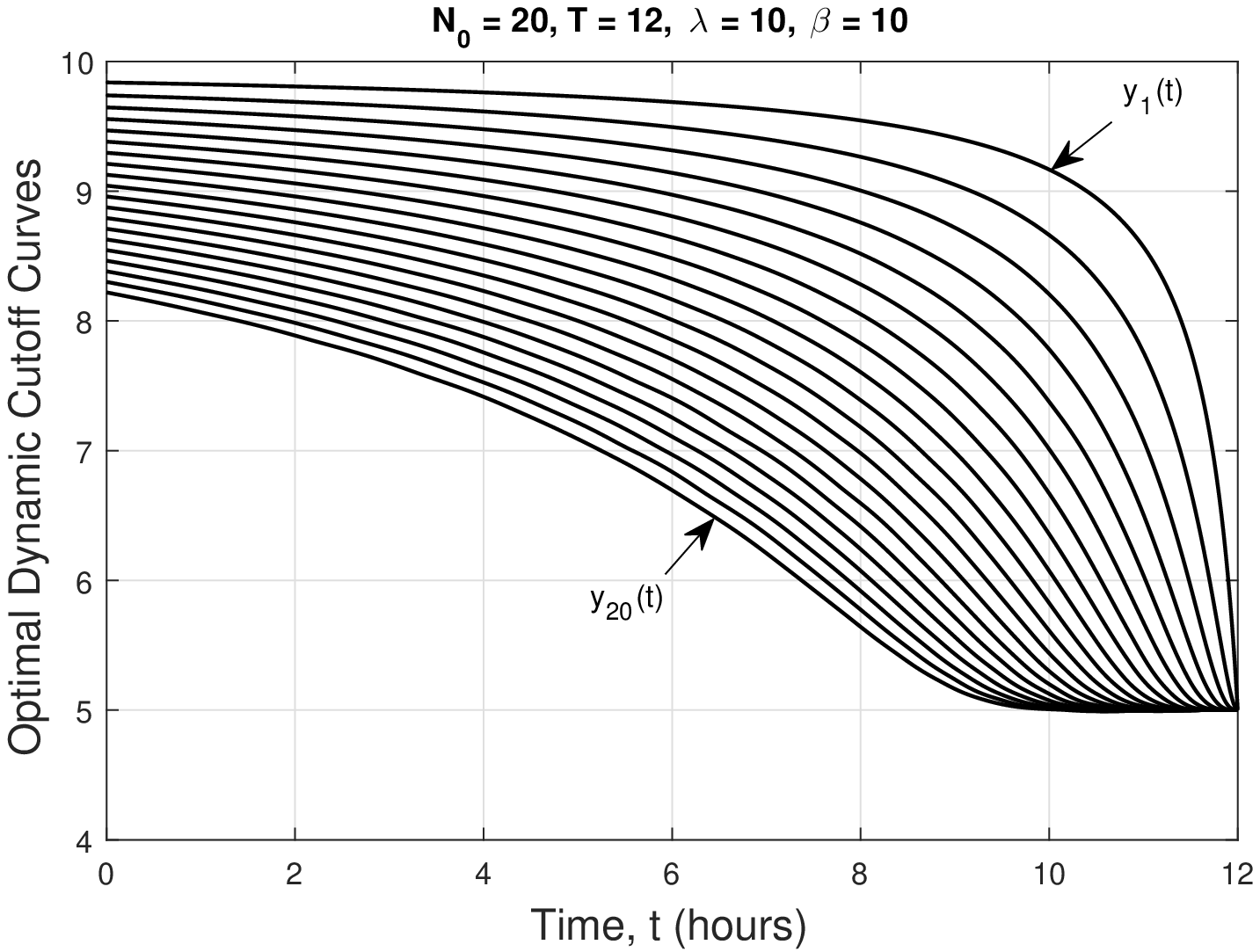} \label{cutoff_unif_fig}}
\caption{Optimal cutoff curves for (a) exponentially distributed and (b) uniformly distributed arrival characteristics.}
\label{cutoff_fig}
\end{figure*}

\begin{figure*}[t!]
	\centering
	\vspace{-0.0in}
	\subfloat[]{\includegraphics[width = 3.1in]{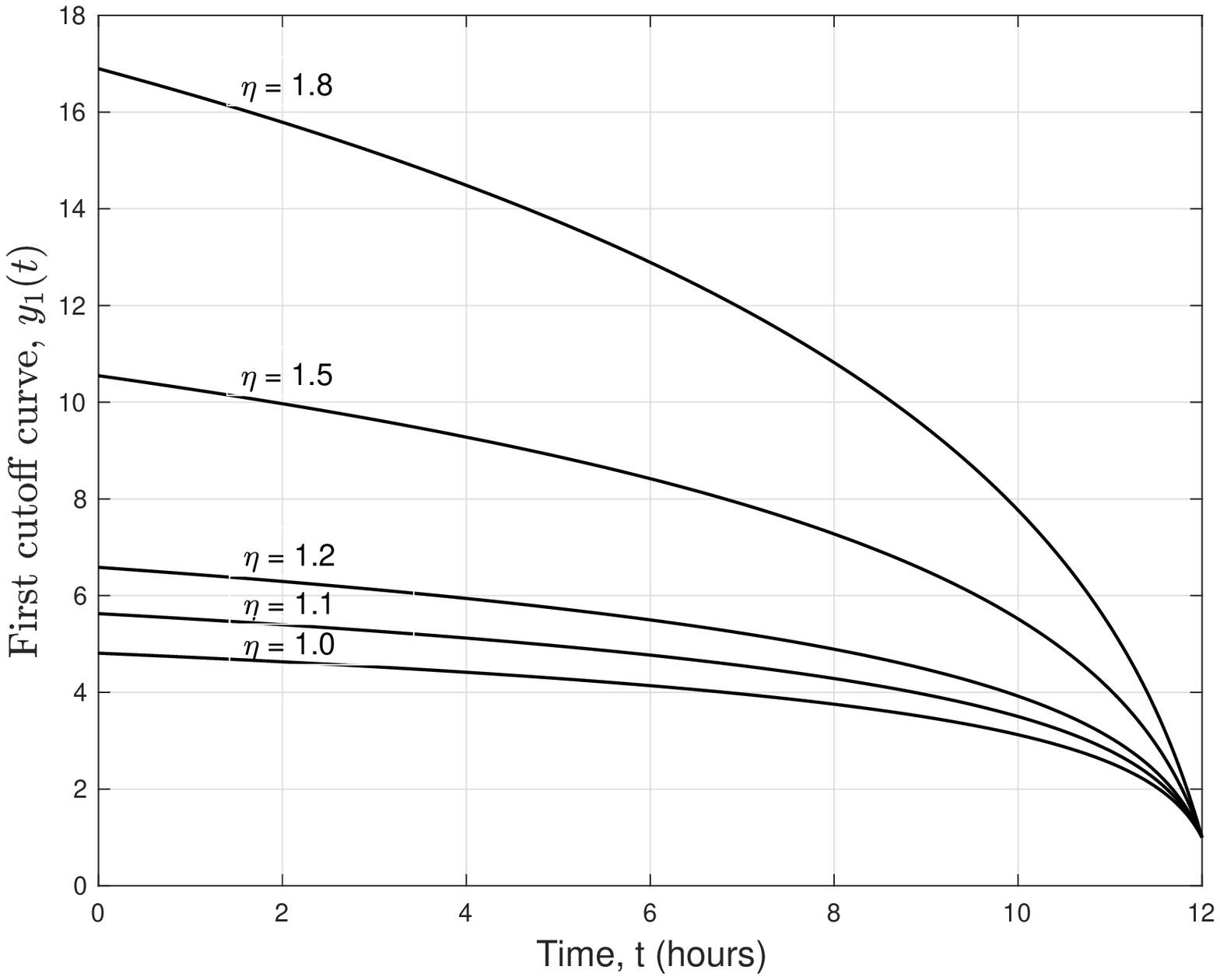} \label{exp_eta_fig}} \ \
	\subfloat[]{\includegraphics[width = 3.1in]{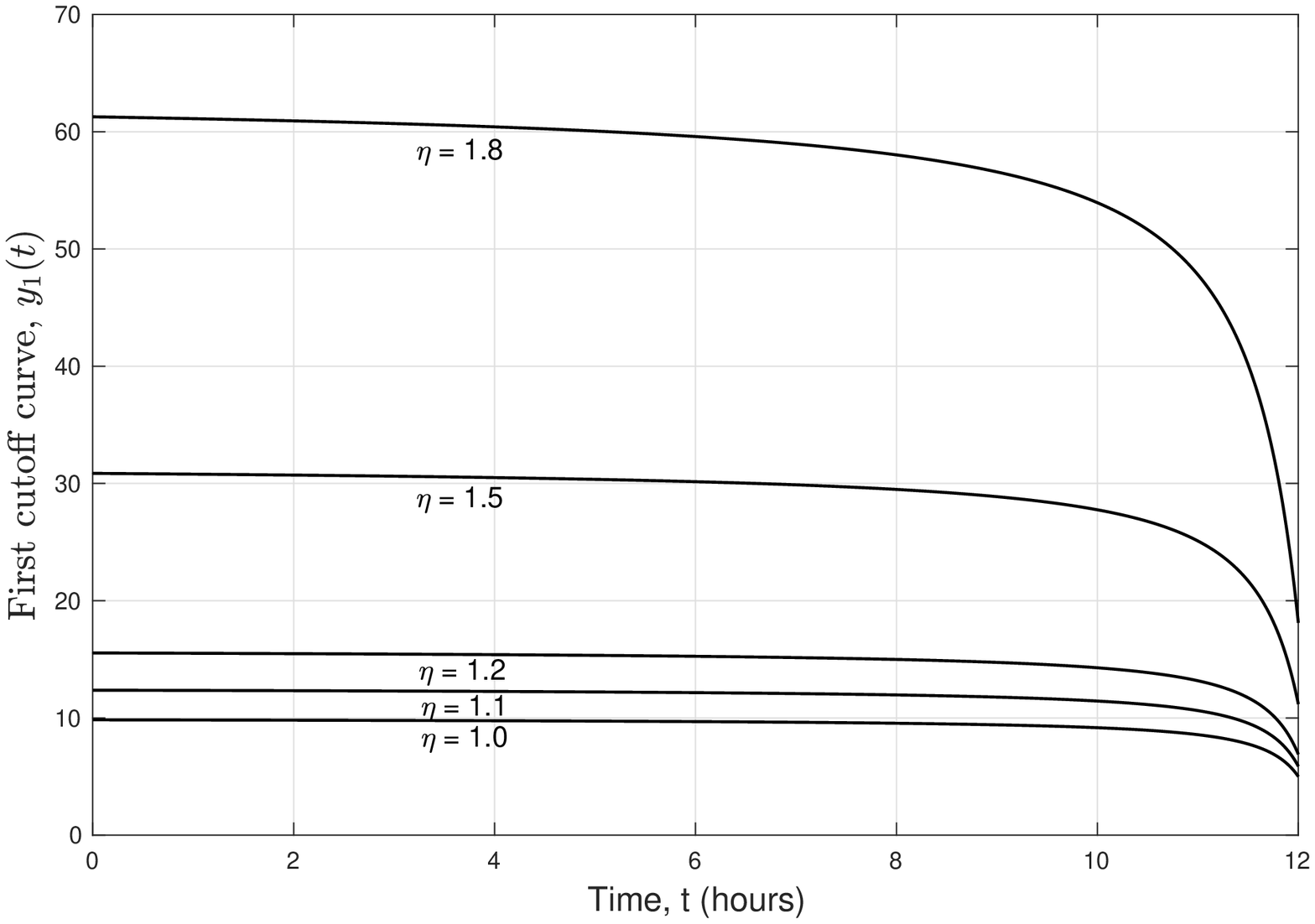} \label{unif_eta_fig}}
	\caption{Optimal first cutoff curves with varying QoE parameter $\eta$ for (a) exponentially distributed and (b) uniformly distributed arrival characteristics.}
	\label{cutoff_fig_eta}
\end{figure*}


\begin{figure*}[t!]
	\vspace{-0.0in}
	\centering
	\subfloat[]{\includegraphics[width = 3.1in]{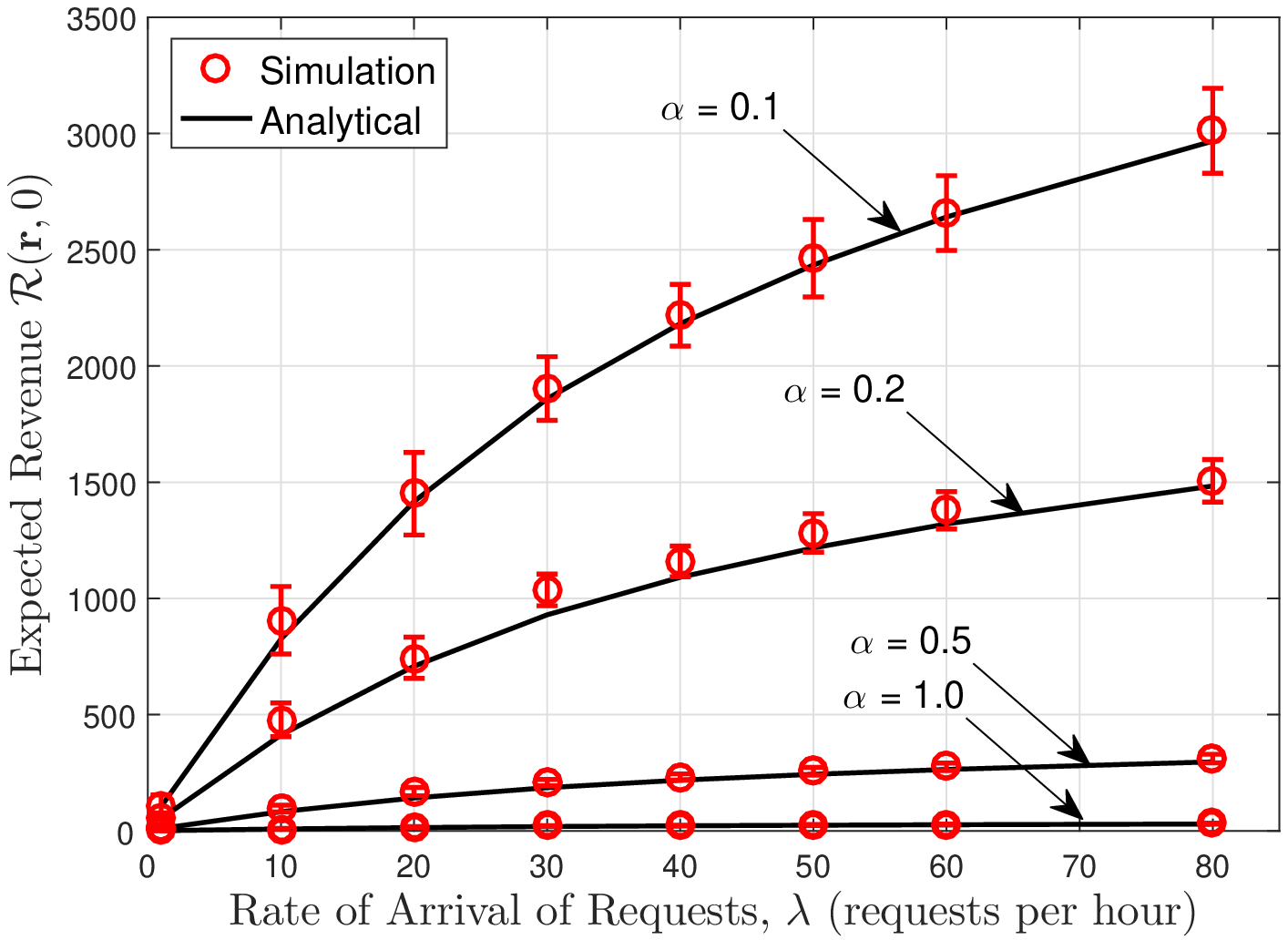} \label{revenue_lambda_exp}} \ \ \ \
	\subfloat[]{\includegraphics[width = 3.1in]{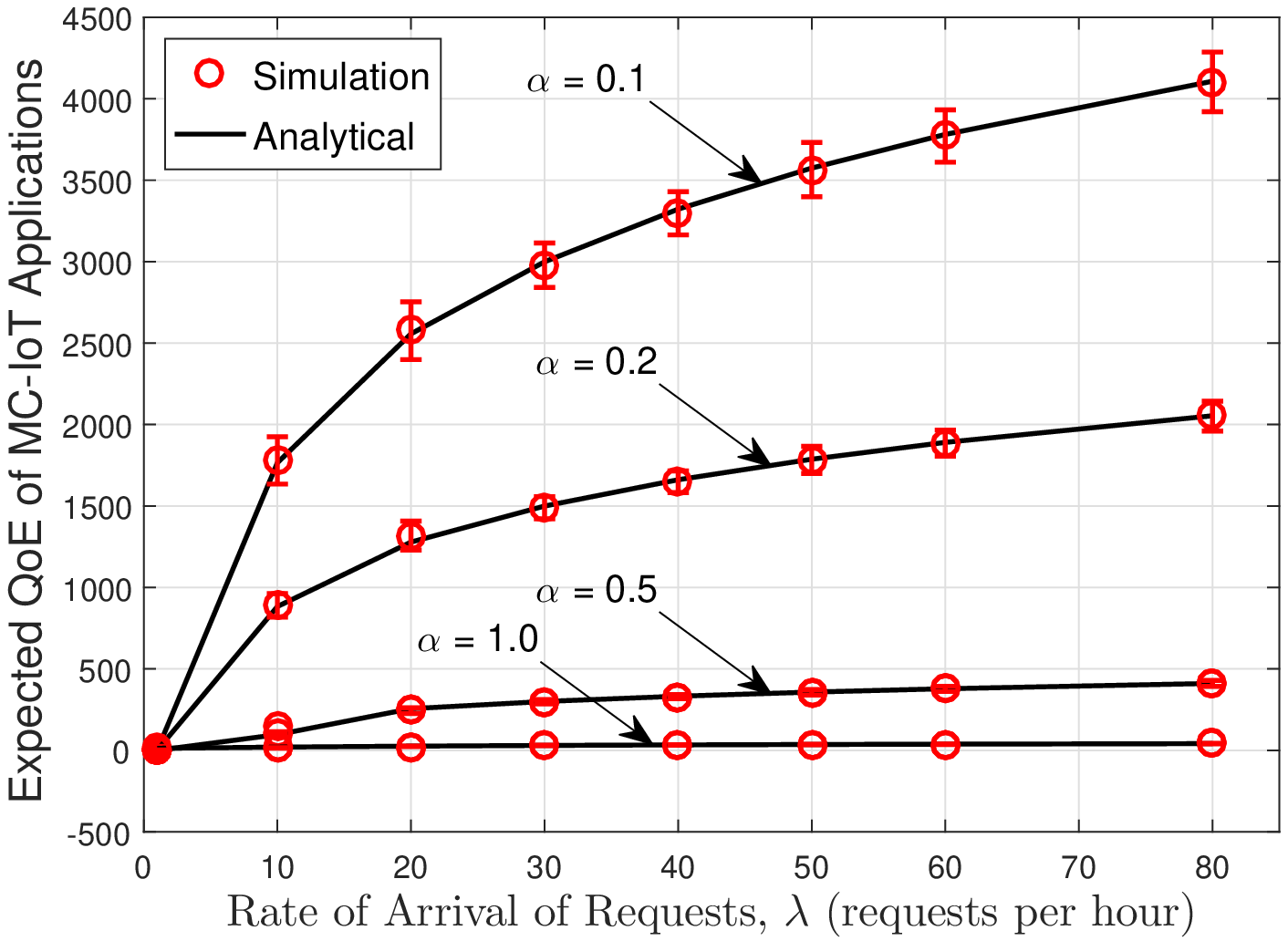} \label{welfare_lambda_exp}}\\
	\subfloat[]{\includegraphics[width = 3.1in]{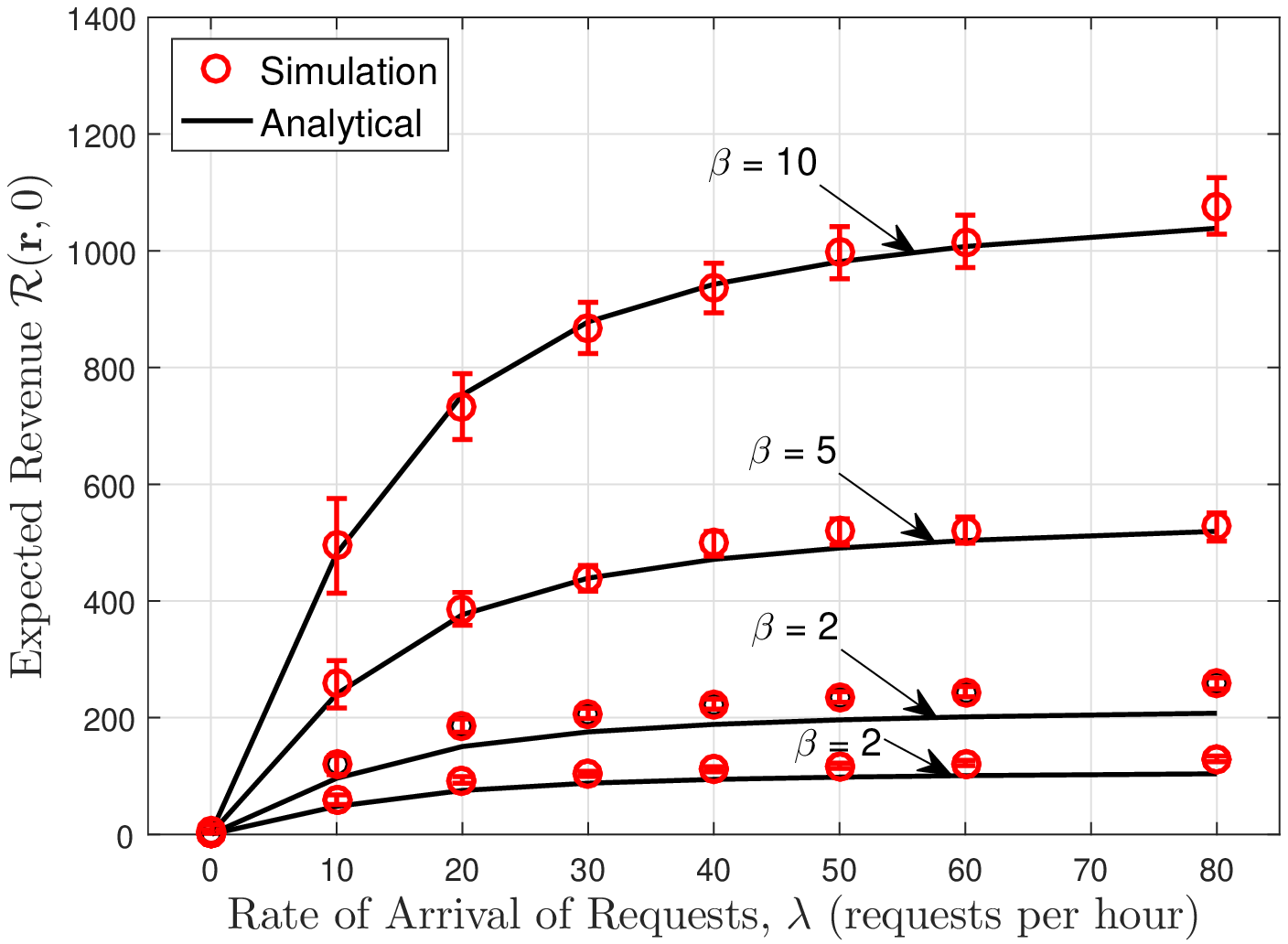} \label{revenue_lambda_unif}} \ \ \ \
	\subfloat[]{\includegraphics[width = 3.1in]{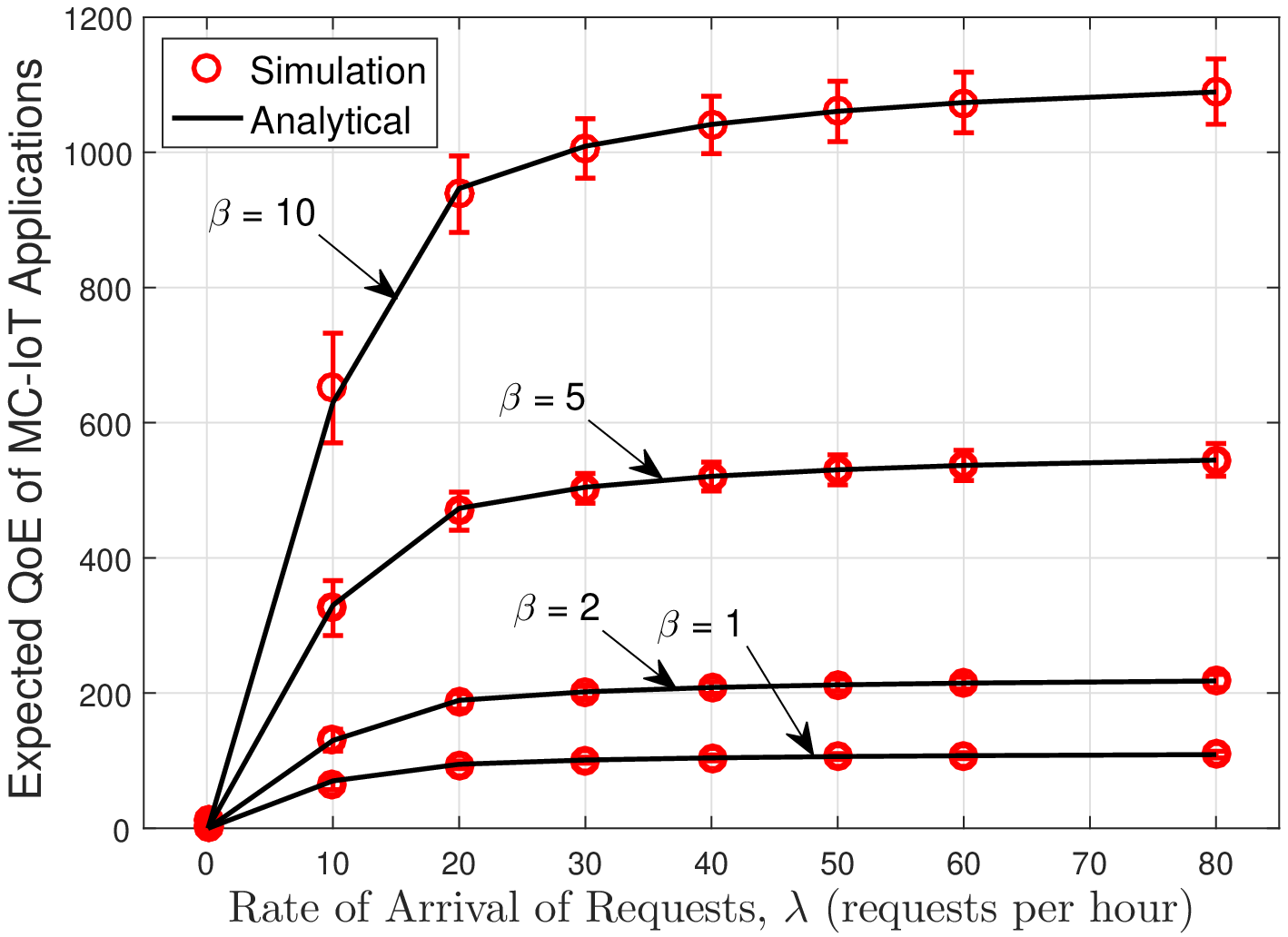} \label{welfare_lambda_unif}}
	\caption{Effect of arrival rate on the expected revenue of the CSP ((a) and (c)) and expected QoE of the users ((b) and (d)) for an exponentially and uniformly distributed arrival characteristic respectively.}
	\label{vs_lambda}
\end{figure*}

 
\vspace{-0.0in}
\section{Numerical Experiments \& Discussion} \label{Sec:Numerical_Experiments}
In this section, we present the results of the numerical experiments performed to evaluate the performance of our proposed allocation and pricing framework.
We first describe the simulation setup and parameter selection. We then provide the results of the simulation and study the behaviour of the framework under varying parameters. Finally, we provide a comparison of the proposed framework with some benchmark allocation strategies and discuss the insights.

\vspace{-0.0in}
\subsection{Experiment Setup}
We assume a CSP with $k=5$ fog data centers in addition to the main cloud servers (Internet data centers) serving a fixed MC-IoT installation area. Each fog data center has $n_i = 20, i = 1, \ldots, 5$, available VMIs for allocation to incoming MC-IoT requests. We have selected a small fog computing architecture in the experiments for the sake of simplicity. However, the experiments can be extended to larger topologies without loss of generality in the results.

The fog data centers are located at increasing distances from the MC-IoT devices which results in increasing latency experienced by the devices while accessing the corresponding data centers. We assume the RTT of the fog nodes experienced by the MC-IoT installation are $l_1 = 0.1$, $l_2 = 0.2$, $l_3 = 0.4$, $l_4 = 0.6$, and $l_5 = 0.8$ ms respectively. The transmission delay over the air interfaces is assumed to be fixed at $\tau^{(o)} = 0.1$ ms. The workload processing times of the VMIs, which depends on the provisioned computing resources, is considered to be uniformly distributed in the interval [0.2,1] ms in the simulations. A demand horizon of $T=12$ hours is assumed during which the allocation takes place. The constant $\eta$ in the QoE function is selected to be $\eta = 1$ for simplicity of results.

\begin{figure*}[t!]
	\centering
	\vspace{-0.0in}
	\subfloat[]{\includegraphics[width = 3.1in]{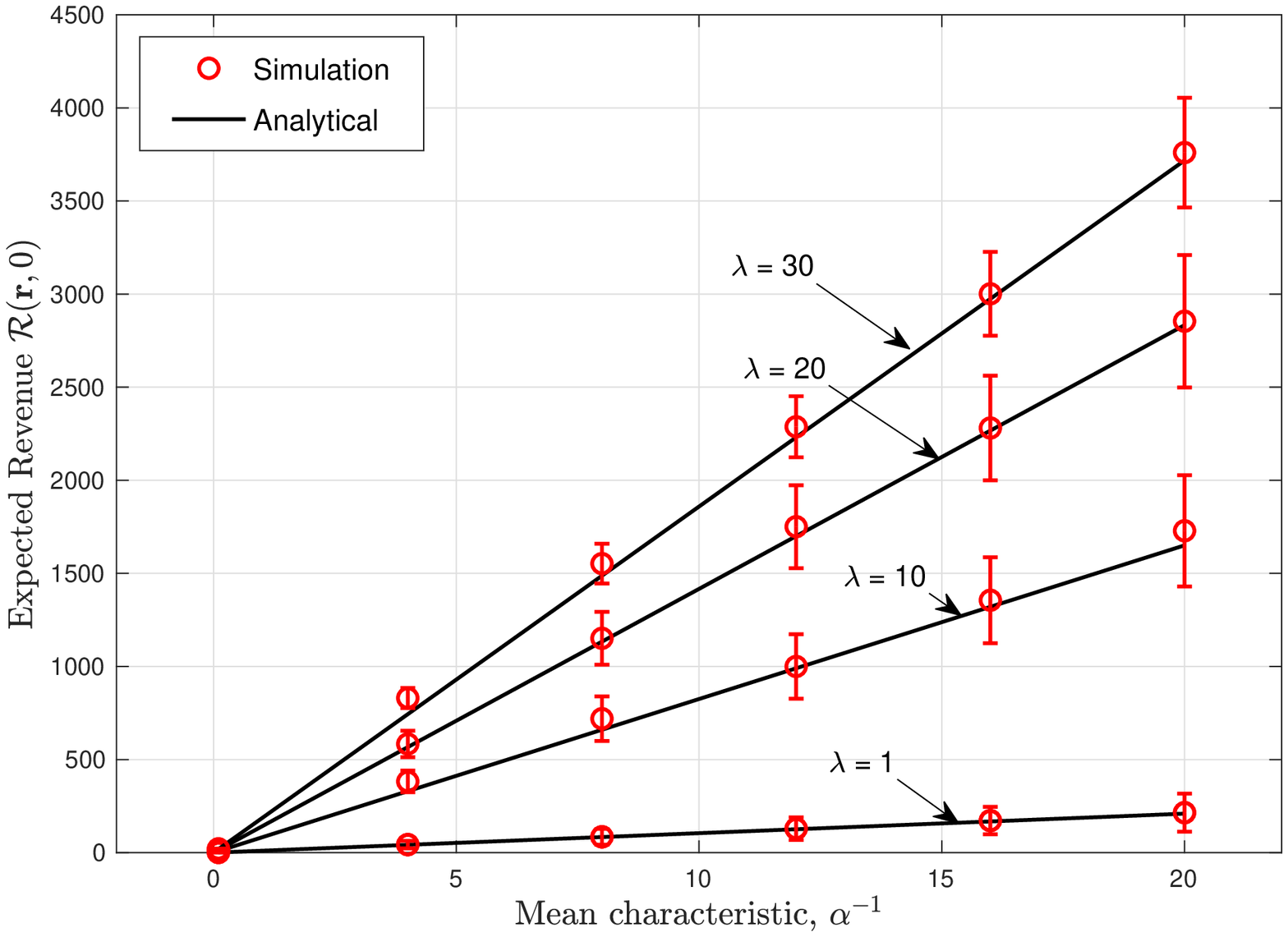} \label{revenue_alpha_exp}} \ \ \ \
	\subfloat[]{\includegraphics[width = 3.1in]{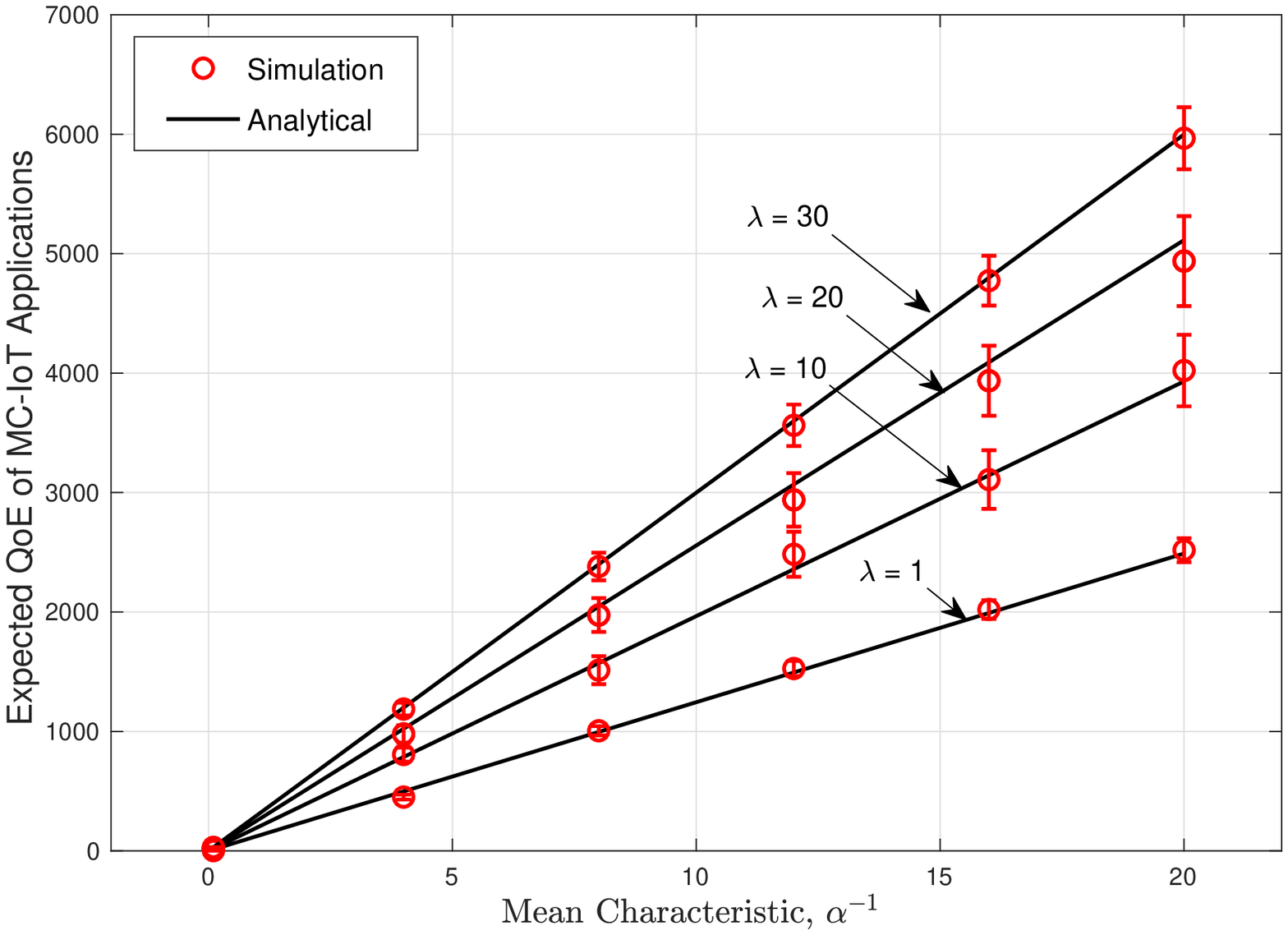} \label{welfare_alpha_exp}}\\
	\subfloat[]{\includegraphics[width = 3.1in]{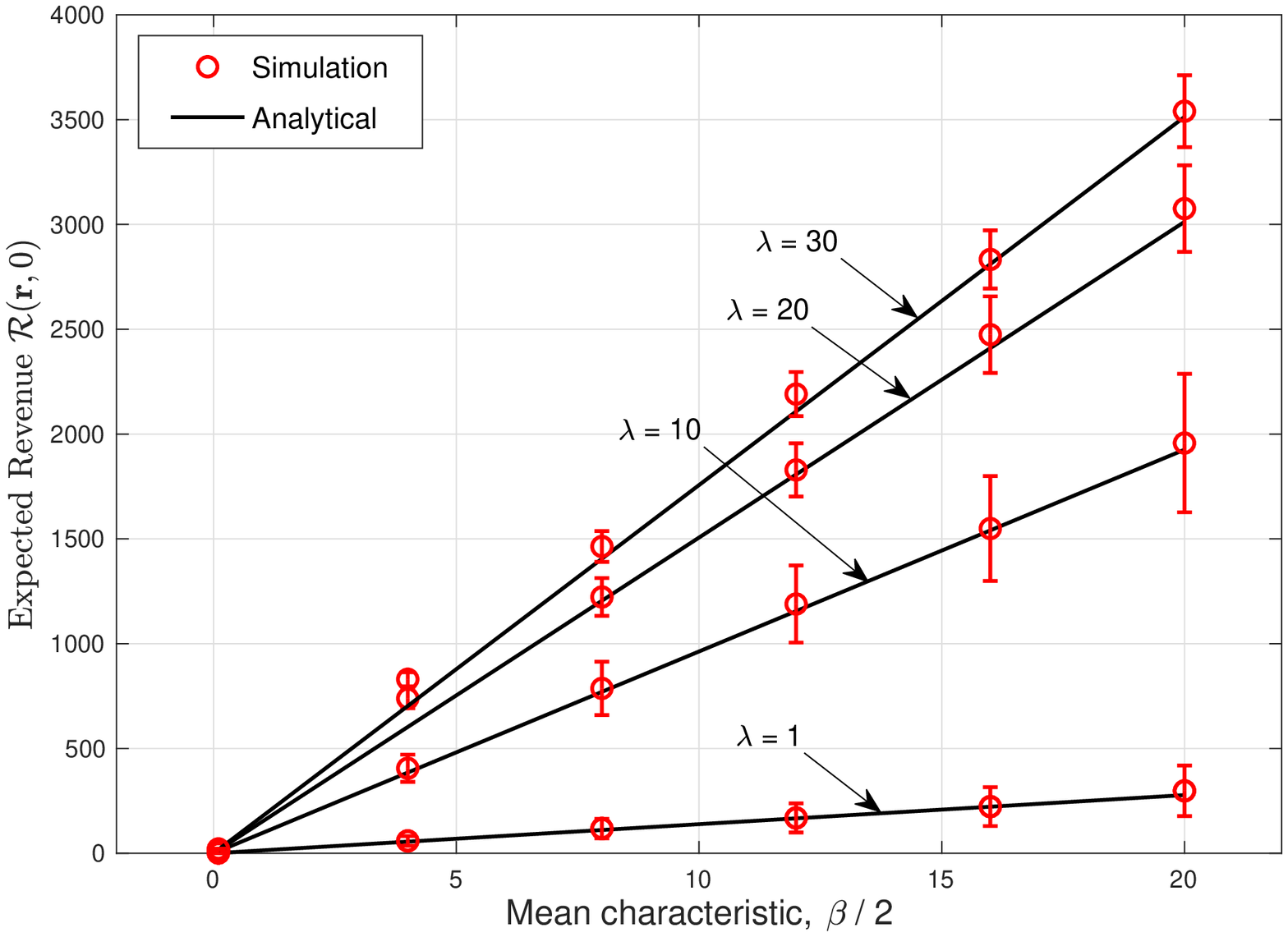} \label{revenue_alpha_unif}} \ \ \ \
	\subfloat[]{\includegraphics[width = 3.1in]{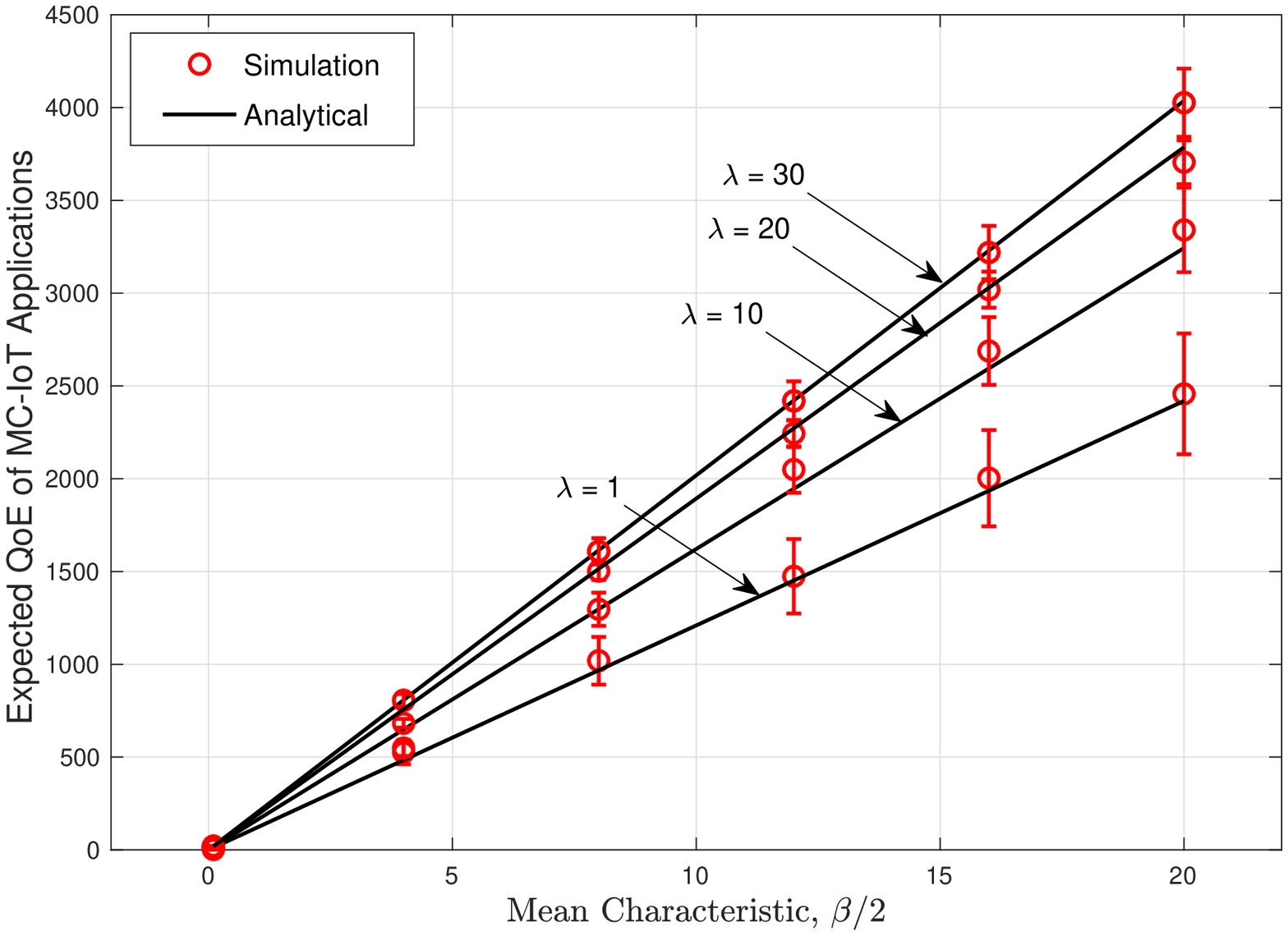} \label{welfare_alpha_unif}}
	\caption{Effect of the mean characteristic on the expected revenue of the CSP ((a) and (c)) and expected QoE of the users ((b) and (d)) for an exponentially and uniformly distributed arrival characteristic respectively. \vspace{-0.0in}}
	\label{vs_alpha}
\end{figure*}

\begin{figure*}[h]
	\centering
	\vspace{-0.0in}
	\subfloat[]{\includegraphics[width = 3.2in]{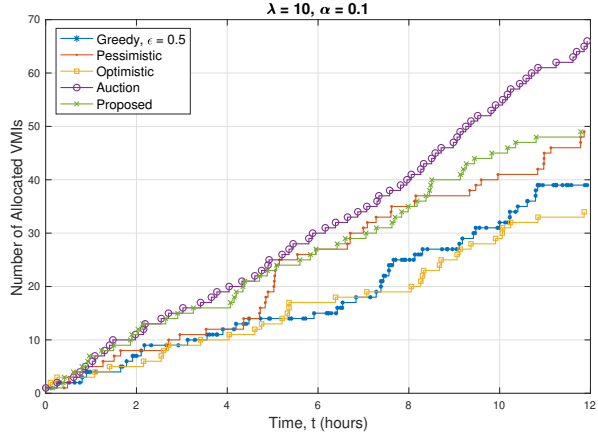} \label{num_allocated}} \ \ \ \ \
	\subfloat[]{\includegraphics[width = 3.2in]{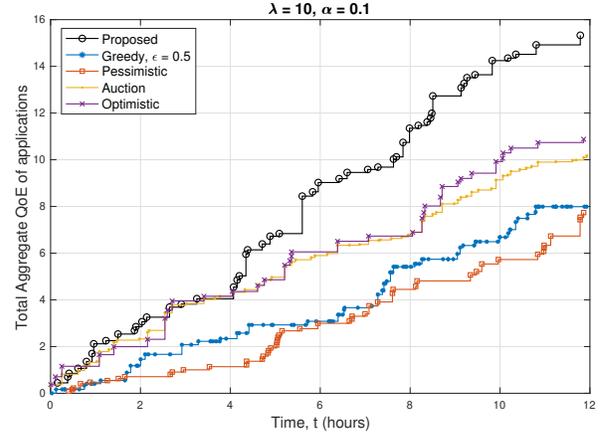} \label{qoe_accumulated}}\\ \vspace{-0.0in}
	\caption{Time evolution of the allocation and total QoE of the users for exponentially distributed characteristic.}	\label{time_evolution}
\end{figure*}


\begin{figure*}[h!]
	\centering
	\vspace{-0.0in}
	\subfloat[]{\includegraphics[width = 3.2in]{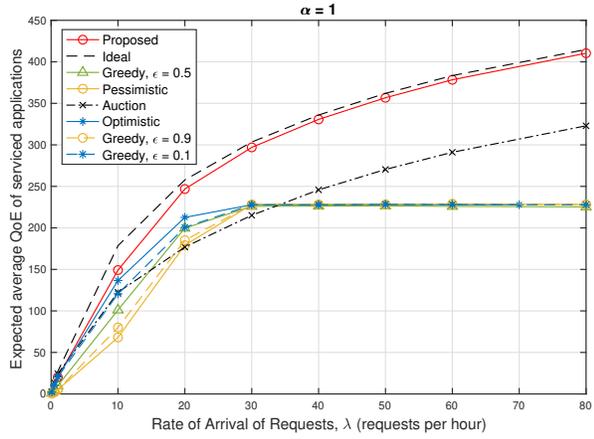} \label{comp_exp_lambda}} \ \ \ \ \
	\subfloat[]{\includegraphics[width = 3.2in]{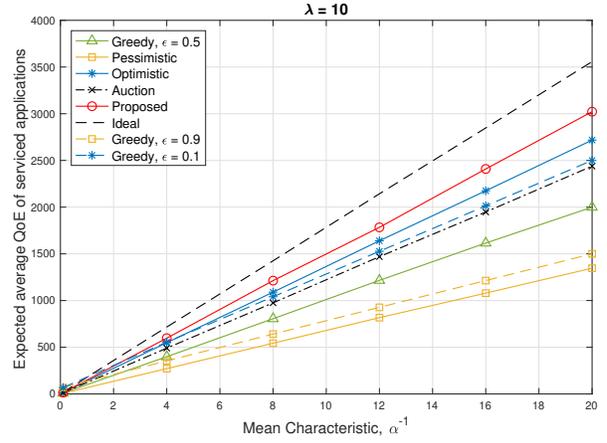} \label{comp_exp_mean}}\\ \vspace{-0.0in}
	\subfloat[]{\includegraphics[width = 3.2in]{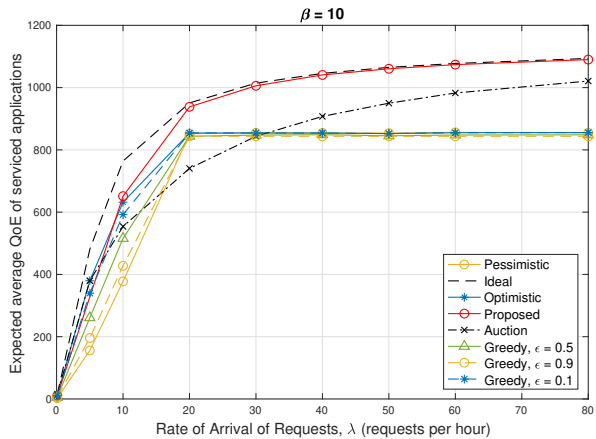} \label{comp_unif_lambda}} \ \ \ \ \
	\subfloat[]{\includegraphics[width = 3.2in]{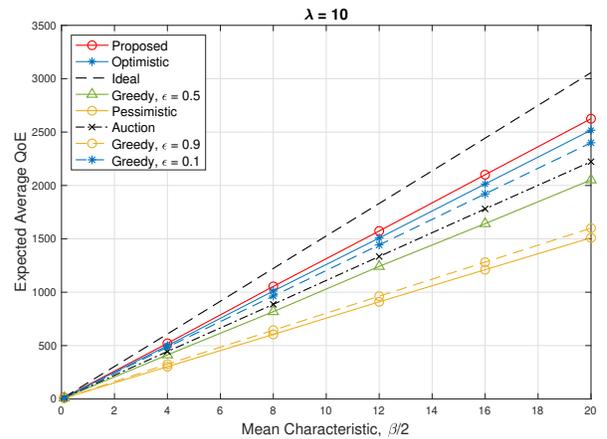} \label{comp_unif_mean}}
	\caption{Comparison of expected QoE for exponentially distributed (a \& b) and uniformly distributed ((c) \& (d)) arrival characteristic.}
	\label{comparison_fig}
\end{figure*}

\begin{figure*}[h!]
	\centering
	\vspace{-0.0in}
	\subfloat[]{\includegraphics[width = 3.2in]{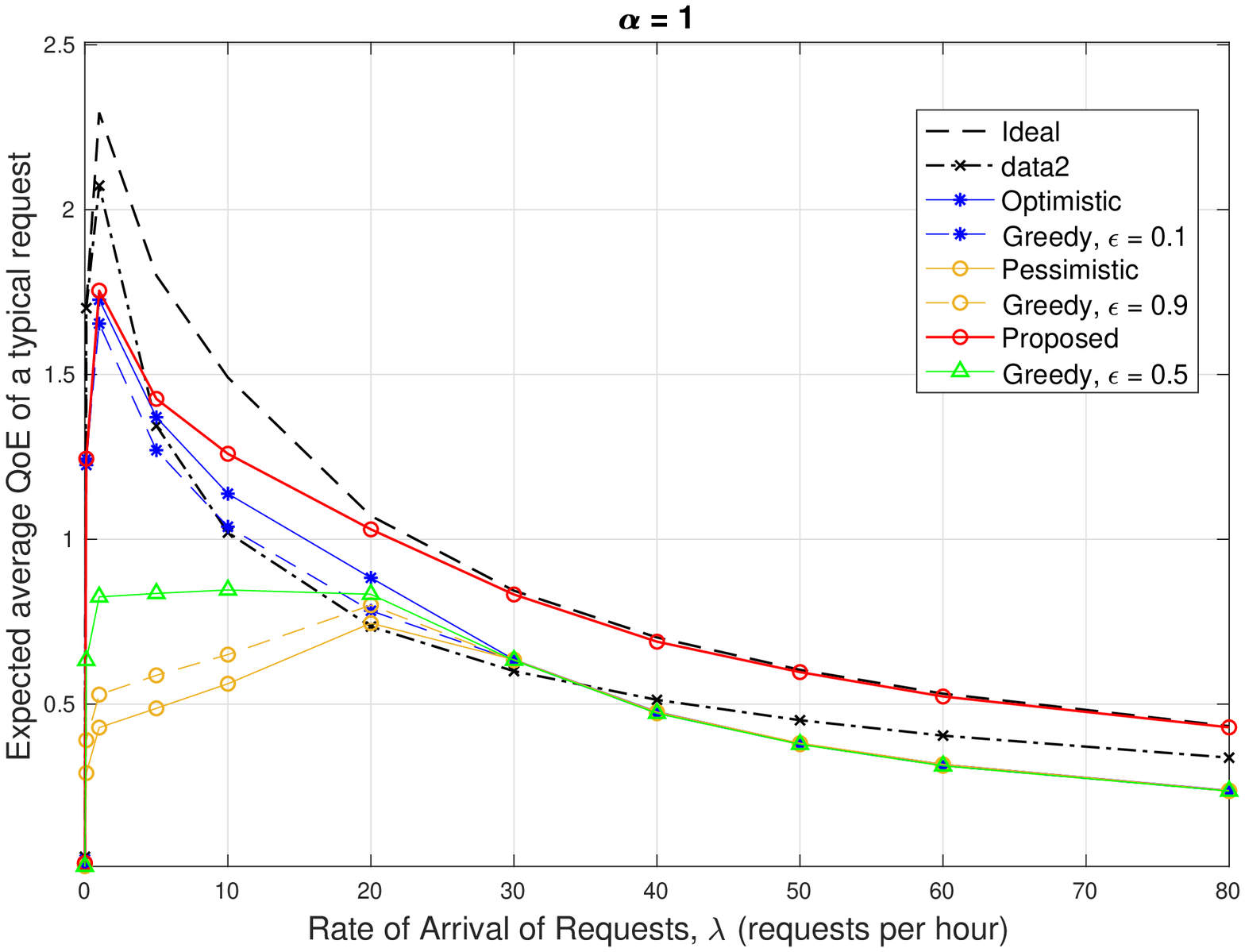} \label{qoe_user_exp}} \ \ \ \ \
	\subfloat[]{\includegraphics[width = 3.2in]{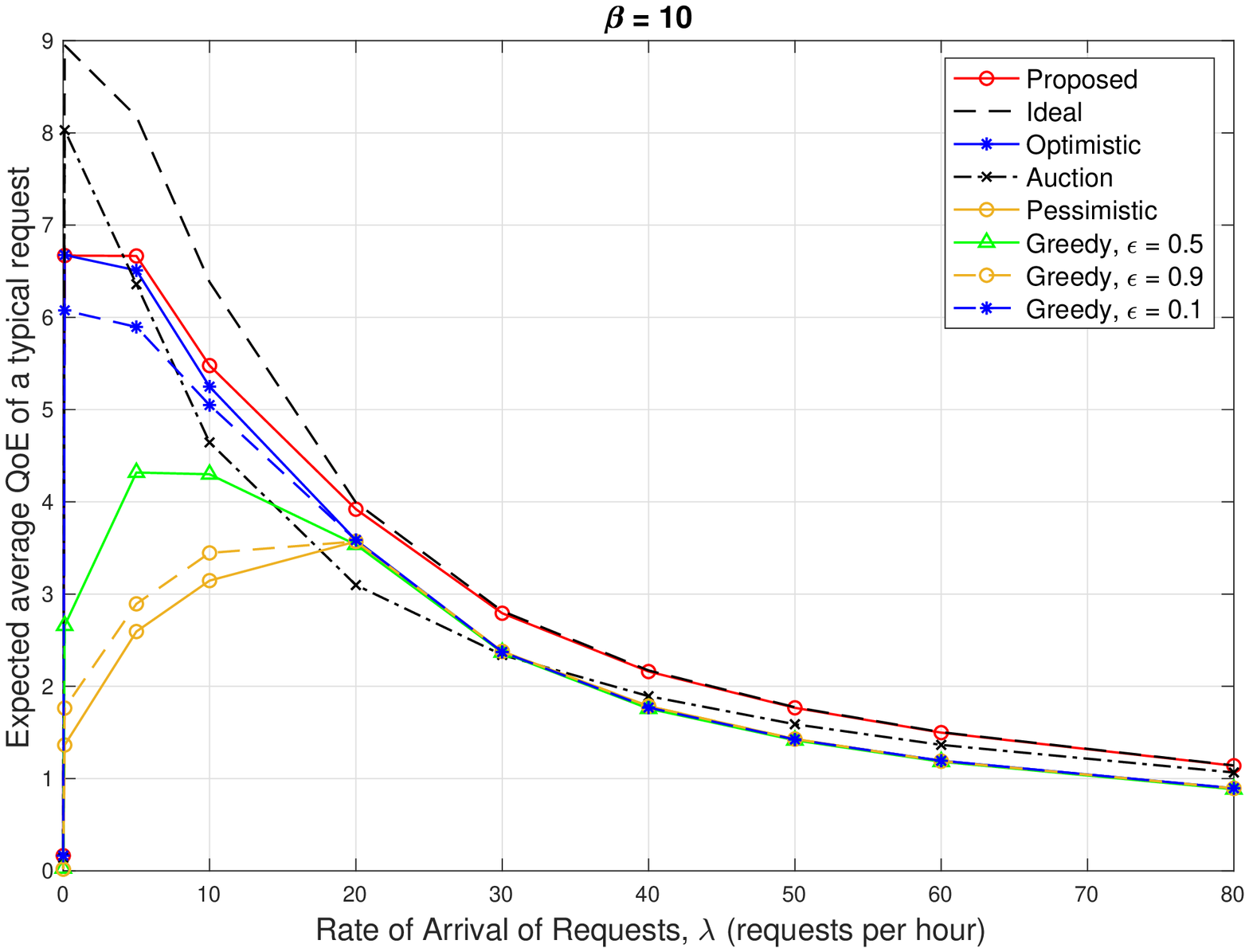} \label{qoe_user_unif}}\\ \vspace{-0.0in}
	\caption{Comparison of average expected QoE for a typical request for (a) exponentially distributed and (b) uniformly distributed arrival characteristic.}
	\label{comparison_user_fig}
\end{figure*}

The MC-IoT applications generate computational requests according to a homogeneous Poisson process that arrive sequentially at the CSP with an arrival rate $\lambda = 10$ requests per hour unless otherwise specified. The reported required response rate of MC-IoT applications is simulated as i.i.d. random variables distributed according to an exponential distribution with mean $\alpha^{-1}$, where $\alpha = 1$, or according to a uniform distribution in the interval $[0, \beta]$, where $\beta = 10$. In the following subsections, we provide the results of the simulations and the obtained insights.
\vspace{-0.0in}

\subsection{Simulation Results}
First we obtain the optimal dynamic cutoff curves obtained for dynamic revenue maximizing VMI allocation. Based on the assumption that the reported required response rate of MC-IoT applications is exponentially distributed and uniformly distributed, the cutoff curves are computed by iteratively solving the differential equations provided in Theorem 2. The first 20 dynamic thresholds for both cases are provided in Fig.~\ref{cutoff_fig}. These are used as a policy for classification of the arriving requests. In general, it can be observed that the thresholds decrease, i.e., the qualification standard is reduced, as the time increases. This implies that the CSP becomes less and less selective as the allocation period is ending to ensure that the maximum number of VMIs are allocated to upcoming requests. However, when there is a lot of time remaining before the allocation period ends, the CSP is more selective ensuring that only the task that reports a sufficiently high required response rate is allocated to a VMI.
It can also be observed from Fig.~\ref{cutoff_exp_fig} that the cutoff curves for the exponentially distributed arrivals have a decreasing allocation margin as the number of VMIs increase. This is due to the possibility of arrival of an highly delay sensitive application for which the barrier is set extremely high. On the other hand, Fig.~\ref{cutoff_unif_fig} shows that under uniformly distributed arrival characteristic, the allocation margin for each of the VMIs is almost constant as the arrival characteristics are equally likely. In both cases, as time approaches the deadline, the cutoff curves approach a constant value equal to the virtual valuation of the agents. It follows that at terminal time, the proposed mechanism is no better than a first-price auction in terms of revenue generation.

In Fig.~\ref{cutoff_fig_eta} illustrates the impact of the QoE parameter $eta$ on the optimal cutoff thresholds. We plot the first cutoff curve, i.e., $y_1(t)$ for varying values of $\eta$ in the case of exponentially distributed and uniformly distributed arrival types in Fig.~\ref{exp_eta_fig} and \ref{unif_eta_fig} respectively. It is clear that the mechanism becomes more selective as $\eta$ increases. However, the thresholds increase much sharply with time for the exponential arrivals as compared to the uniform arrivals.

Next, we investigate the behaviour of the proposed allocation and pricing scheme in response to the changing rate of arrival of MC-IoT requests and the mean arrival characteristic for both exponentially and uniformly distributed arrival types. Fig.~\ref{vs_lambda} shows the expected revenue of the CSP and the expected QoE of the users as the rate of arrival or requests increases. It can be observed from Fig.~\ref{revenue_lambda_exp} and Fig.~\ref{revenue_lambda_unif} that the expected revenue of the CSP increases as the arrival rate increases but saturates at high arrival rates as the available VMIs are exhausted. The exponential arrival characteristic results in a higher expected revenue in general due to the possibility of arrival of highly demanding and highly paying requests, which is not the case in the uniform case. Fig.~\ref{welfare_lambda_exp} and Fig.~\ref{welfare_lambda_unif} depict a similar behaviour in the expected QoE of the users in the two cases. Fig.~\ref{vs_alpha} investigates the behaviour of the expected revenue of the CSP and the expected QoE of the users in response to a change in the mean of the arrival characteristic. It is observed from Fig.~\ref{revenue_alpha_exp} and Fig.~\ref{revenue_alpha_unif} that the expected revenue increases linearly with the mean of the arrival characteristic for both exponential and uniform arrival types. This is because a higher arrival type raises the qualification standard for allocation and consequently the prices. Hence, it does not saturate as the arrival type or rate increases. Finally, Fig.~\ref{welfare_alpha_exp} and Fig.~\ref{welfare_alpha_unif} show a similar increasing behaviour in the expected QoE of the users with an increase in the mean arrival characteristic. 

\vspace{-0.0in}
\subsection{Comparison}
To illustrate the performance of the developed optimal allocation mechanism in terms of the QoE of MC-IoT applications, we compare our proposed allocation scheme with the following benchmark strategies:
\begin{enumerate}
\item \textbf{Ideal Allocation:} In this case, we assume there is no uncertainty in the future arrivals and the delay tolerances of all the requests are known a priori. This case is similar to the advance scheduling or reservation scenario where the CSP is aware of the delay tolerances of the applications in advance. It is a useful benchmark since it provides the theoretical upper bound to the achievable QoE under any possible allocation strategy. Note that a loss due to sequentiality and incomplete information will always be there with reference to the ideal allocation.
\item \textbf{Pessimistic Allocation:} In this strategy, the CSP uses the threshold based allocation policy, however, it adopts a pessimistic approach towards future arrivals, i.e., assumes that more delay tolerant requests will arrive in the future as compared to the current one, and allocates the best available VMIs to the requesting applications first. This strategy does not employ any foresight and only makes myopic decisions based on the current arriving task. Hence it attempts to maximize the QoE and the generated revenue by allocating the best available resources to incoming tasks first.
\item \textbf{Optimistic Allocation:} The optimistic allocation policy is the opposite of the pessimistic allocation policy. It always assumes that the arrivals in the future will be more QoE sensitive than the current task at hand. Hence, it saves the best available VMIs while allocating the VMI with the highest end-to-end delay to the current request expecting more QoE sensitive requests to arrive in the future.
\item \textbf{$\epsilon$-Greedy Allocation:} In this strategy, the CSP allocates one of the available VMI in any of the fog nodes with a probability $\epsilon$ to a requesting application with a probability $\epsilon$. In other words, it has an expectation about the nature of the requests that will arrive in the future and hence allocates the available VMIs to an upcoming request with $\epsilon$ probability.
\item \textbf{Periodic Auction:} In this strategy, the available VMIs are periodically auctioned to the requests. The delay tolerances of incoming requests are considered as bids in the auction and the CSP periodically collects the bids and selects the requests to allocate the VMIs. We use a first price auction strategy whereby the highest bidder wins and receives the best available VMI. We assume that the VMIs are allocated one unit at a time. This strategy is adapted from the work in~\cite{most_related_work_comparison}.
\end{enumerate}

Finally, the proposed optimal dynamic strategy makes use of the statistical information about the future arriving tasks to make dynamically optimal decisions for allocating VMIs to arriving requests strategically. For the scenario considered in the simulations, we compare the proposed approach with each of the above mentioned approaches in terms of the average QoE experienced by the MC-IoT applications. 
We first provide a time evolution of the allocation procedure in Fig.~\ref{time_evolution}. Fig.~\ref{num_allocated} shows the progression of the number of VMIs allocated as time goes on for each allocation strategy. It can be observed that the auction procedure is able to allocate the maximum number of VMIs while the optimistic approach is able to allocate the least number of VMIs. Fig.~\ref{qoe_accumulated} shoes the corresponding progression of total aggregate QoE with time. Note that while our proposed mechanism does not result in allocation of the most VMIs, but it leads to the maximum aggregate QoE among all the other approaches. Similarly, while auction does allocate more VMIs, it is able to achieve an aggregate QoE similar to the optimistic approach.

The results of the average expected QoE achieved by serviced applications in comparison to the benchmark schemes are provided in Fig.~\ref{comparison_fig}. Fig.~\ref{comp_exp_lambda} and Fig.~\ref{comp_exp_mean} illustrate the average QoE against varying the rate and mean characteristic of the arrival respectively for an exponentially distributed required response rate. Fig.~\ref{comp_unif_lambda} and Fig.~\ref{comp_unif_mean} illustrate the average QoE of serviced applications against varying the rate and mean characteristic of the arrival respectively for a uniformly distributed required response rate. It is clear from the results that the proposed allocation strategy provides the best expected QoE to the users as well as maximizes the revenue of the CSP. However it does suffer from loss of sequentiality as compared with the Ideal allocation scheme. Furthermore, random pairings result in a very low expected QoE while the greedy and optimistic approaches lead to an intermediate average QoE of the users. Another striking result is that the auction strategy may not always be better than other benchmark strategies. Under low arrival rates, the optimistic, greedy, and pessimistic may outperform the auction based strategy. \textcolor{black}{Fig.~\ref{comparison_user_fig} shows the comparison of the average QoE of a typical application request against varying arrival rate of requests for both exponentially (Fig.~\ref{qoe_user_exp}) and uniformly distributed (Fig.~\ref{qoe_user_unif}) arrival characteristic. In general, when the rate of arrival of requests is low, there is a lower expected average QoE for a typical request since the total number of highly demanding applications is lower. This increases sharply as the rate of arrival increases. However, when the arrival rate is very high, the average QoE for a typical application declines since only a few of them can be serviced due to limited resource availability.}

\vspace{-0.0in}
\section{Conclusions} \label{Sec:Conclusion}
In this paper, we provide a QoE based revenue maximizing dynamic resource allocation and pricing framework for fog-enabled MC-IoT applications. We propose an implementable mechanism for allocation of  different VMIs available at the fog nodes that result in varying end-to-end delay for user applications.
As opposed to existing works in the literature that focus on static pairing of computational tasks and available fog resources, the proposed resource allocation and pricing strategy is dynamic with instantaneous decision-making as well as takes the hierarchical fog-cloud architecture into account. The developed framework provides an optimal threshold based classification mechanism that uses statistical information about the MC-IoT requests arriving in the future to make dynamically optimal real-time allocation decisions resulting in maximizing revenue of the CSP, based on the QoE of the users. The dynamic thresholds can be pre-computed and used as a lookup table for real-time decision making. Numerical results confirm that the proposed allocation scheme significantly performs better in terms of the QoE achieved by the users in comparison with other benchmark allocation schemes.

\appendices

\section{Proof of Theorem~\ref{single_theorem}} \label{proof_single_theorem}

\textcolor{blue}{}
The proof for the dynamically optimal revenue maximizing curves requires the characterization of the expected revenue of the CSP over the allocation period is a dynamic threshold is used to filter incoming requests. This construction of the proofs is based on the works of Gerkshov and Moldovanu~\cite{revenue_maximization}.
Consider the case where at time $t$ only a single VMI with response rate $r_1$ is available to the CSP for allocation up to time $T$. The expected revenue with a single cutoff curve $y_1(t)$ can be expressed as follows:
\begin{align}
\mathcal{R}(\{r_1\},t) = r_1^{\frac{1}{\eta}} \int_t^T y_{1}^{\frac{1}{\eta}}(s) h_{1}(s) ds,
\end{align}
where $h_{1}(s)$ is the probability density of waiting time until the first arrival of a request with a required response rate of greater than $y_{1}(s)$. The density can be represented by the first arrival in a non-homogeneous Poisson process with intensity $\lambda (1 - F_X(y_{1}(s)))$. For the sake of analytical tractability, we let $\hat{X} = X^{\frac{1}{\eta}}$. Therefore we will use the intensity $\lambda (1 - F_{\hat{X}}(y_{1}^{\frac{1}{\eta}}(s)))$ instead of the former one.
Note that the density of homogeneous Poisson task arrival process is thinned by a factor $(1 - F_{\hat{X}}(y_1^{\frac{1}{\eta}}(s)))$, which represents the probability that the performance improvement of the arriving task is above the set qualification threshold $y_1^{\frac{1}{\eta}}(s)$. The density can be expressed as follows~\cite{stochastic_book}:
\begin{align}
&h_{1}(s) = \lambda  (1 - F_{\hat{X}}(y_{1}^{\frac{1}{\eta}}(s)))  \times \notag \\ &  \exp \left(- \int_t^s \lambda (1 - F_{\hat{X}}(y_{1}^{\frac{1}{\eta}}(u))) du \right) ,   \ \ t \leq s \leq T.
\end{align}
Let $H(s) = \int_t^s \lambda \left( 1 - F_{\hat{X}}(y_{1}^{\frac{1}{\eta}}(u)) \right) du$. The expected revenue can then be written as follows:
\begin{align} \label{expected_revenue}
\mathcal{R}(\{r_1\},t) = r_1^{\frac{1}{\eta}} \int_t^T  F_{\hat{X}}^{-1} \left(  1 - \frac{H^{\prime}(s)}{\lambda}  \right)  H^{\prime} (s) e^{-H(s)} ds.
\end{align}
The kernel of integration can be expressed as follows:
\begin{align}
L(s,H(s), H^{\prime}(s)) =  F_{\hat{X}}^{-1} \left(  1 - \frac{H^{\prime}(s)}{\lambda}  \right) H^{\prime} (s) e^{-H(s)}.
\end{align}
In order to maximize the expected revenue in~\eqref{expected_revenue}, we employ the Euler-Lagrange equation from the calculus of variations~\cite{calculus_variations}. The necessary condition for the revenue maximizing cutoff curves can be expressed as $\frac{\partial L(s,H(s), H^{\prime}(s))}{\partial H(s)} - \frac{d}{dt} \frac{\partial L(s,H(s), H^{\prime}(s))}{\partial H^{\prime}(s)} = 0$. The partial derivatives can be expressed as follows:
\begin{align}
\frac{\partial L}{\partial H(s)} = -  F_{\hat{X}}^{-1} \left(  1 - \frac{H^{\prime}(s)}{\lambda}  \right) H^{\prime} (s) e^{-H(s)}.
\end{align}
\begin{align}
\frac{\partial L}{\partial H^{\prime}(s)} =  e^{-H(s)} \left(  F_{\hat{X}}^{-1}  \left(1 - \frac{H^{\prime}(s)}{\lambda}\right) \right. \notag \\ \left. -  \left( \frac{H^{\prime}(s) }{\lambda f_{\hat{X}} \left(F_{\hat{X}}^{-1}  \left(1 - \frac{H^{\prime}(s)}{\lambda}\right)\right)}\right) \right).
\end{align}
\begin{align}
\frac{d}{dt}\frac{\partial L}{\partial H^{\prime}(s)} &=  - \left(  F_{\hat{X}}^{-1}  \left(1 - \frac{H^{\prime}(s)}{\lambda}\right) -    \right. \notag \\ & \left.  \left( \frac{H^{\prime}(s) }{\lambda f_{\hat{X}} \left(F_{\hat{X}}^{-1}  \left(1 - \frac{H^{\prime}(s)}{\lambda}\right)\right)}\right) \right) e^{-H(s)} H^{\prime}(s) + \notag \\ &e^{-H(s)} \left( \frac{- 2 H^{\prime \prime}(s)}{\lambda f_{\hat{X}} \left(F_{\hat{X}}^{-1}  \left(1 - \frac{H^{\prime}(s)}{\lambda}\right)\right)}\right. \notag \\
&\left. - \frac{f_{\hat{X}}^{\prime}\left(F_{\hat{X}}^{-1}  \left(1 - \frac{H^{\prime}(s)}{\lambda}\right)\right) H^{\prime}(s) H^{\prime \prime}(s)}{ \lambda^2 f_{\hat{X}}^3 \left( F_{\hat{X}}^{-1}  \left(1 - \frac{H^{\prime}(s)}{\lambda}\right) \right)} \right),
\end{align}
The Euler-Lagrange equation can be written as follows:
\begin{align}
-  (H^{\prime}(s))^2
+& 2 H^{\prime \prime}(s) + \notag \\
&\frac{f_{\hat{X}}^{\prime}\left(F_{\hat{X}}^{-1}  \left(1 - \frac{H^{\prime}(s)}{\lambda}\right)  \right) H^{\prime}(s) H^{\prime \prime}(s)}{ \lambda f_{\hat{X}}^2 \left( F_{\hat{X}}^{-1}  \left(1 - \frac{H^{\prime}(s)}{\lambda}\right) \right)} = 0.
\end{align}
Then, plugging back $H(s) = \int_t^s \lambda (1 - F_{\hat{X}}(y_1^{\frac{1}{\eta}}(u))) du$, we obtain the following:
\begin{align}
-\lambda (1 - F_{\hat{X}}(y_1^{\frac{1}{\eta}}(s)))^2 - \frac{2}{\eta} F_{\hat{X}}(y_1^{\frac{1}{\eta}}(s)) y_1^{\frac{1}{\eta}-1}(s) y_1^{\prime}(s) - \notag \\
\frac{ f_{\hat{X}}^{\prime}(y_1^{\frac{1}{\eta}}(s)) (1 - F_{\hat{X}}(y_1^{\frac{1}{\eta}}(s))) y_1^{\frac{1}{\eta}-1}(s) y_1^{\prime}(s)}{ \eta f_{\hat{X}} (y_1(s))} = 0.
\end{align}
This can be further expressed as follows:
\begin{align}
\frac{d}{ds}\left(y_1^{\frac{1}{\eta}}(s)\right)  + \frac{\lambda (1 - F_{\hat{X}}(y_1^{\frac{1}{\eta}}(s)))^2}{f_{\hat{X}}(y_1^{\frac{1}{\eta}}(s))}  = \frac{d}{ds} \left( \frac{  1 - F_{\hat{X}}(y_1^{\frac{1}{\eta}}(s))}{  f_{\hat{X}} (y_1^{\frac{1}{\eta}}(s))} \right).
\end{align}
Integrating both sides from $t$ to $T$ results in the following:
\begin{align}
&y_1^{\frac{1}{\eta}}(T) - y_1^{\frac{1}{\eta}}(t) + \lambda \int_t^T  \frac{(1 - F_{\hat{X}}(y_1^{\frac{1}{\eta}}(s)))^2}{f_{\hat{X}}(y_1^{\frac{1}{\eta}}(s))} ds  =  \notag \\ &\left( \frac{  1 - F_{\hat{X}}(y_1^{\frac{1}{\eta}}(T))}{  f_{\hat{X}} (y_1^{\frac{1}{\eta}}(T))} \right) - \left( \frac{  1 - F_{\hat{X}}(y_1^{\frac{1}{\eta}}(t))}{  f_{\hat{X}} (y_1^{\frac{1}{\eta}}(t))} \right)
\end{align}
Using the boundary condition $y_1^{\frac{1}{\eta}}(T) - \frac{1 - F_{\hat{X}}(y_1^{\frac{1}{\eta}}(T))}{f_{\hat{X}}(y_1^{\frac{1}{\eta}}(T))} = 0$, we reach the following expression:
\begin{align}
- y_1^{\frac{1}{\eta}}(t) + \lambda \int_t^T &  \frac{(1 - F_{\hat{X}}(y_1^{\frac{1}{\eta}}(s)))^2}{f_{\hat{X}}(y_1^{\frac{1}{\eta}}(s))}  ds = \notag \\ &- \left( \frac{  1 - F_{\hat{X}}(y_1^{\frac{1}{\eta}}(t))}{  f_{\hat{X}} (y_1^{\frac{1}{\eta}}(t))} \right)
\end{align}
Rearranging the terms results in the following:
\begin{align}
y_1^{\frac{1}{\eta}}(t) =  \frac{  1 - F_{\hat{X}}(y_1^{\frac{1}{\eta}}(t))}{  f_{\hat{X}} (y_1^{\frac{1}{\eta}}(t))}  +  \lambda \int_t^T  \frac{(1 - F_{\hat{X}}(y_1^{\frac{1}{\eta}}(s)))^2}{f_{\hat{X}}(y_1^{\frac{1}{\eta}}(s))} ds. \label{y1}
\end{align}
Equivalently, it can be written as follows:
\begin{align}
y_1(t) =  \left( \frac{  1 - F_{\hat{X}}(y_1^{\frac{1}{\eta}}(t))}{  f_{\hat{X}} (y_1^{\frac{1}{\eta}}(t))}  +  \lambda \int_t^T  \frac{(1 - F_{\hat{X}}(y_1^{\frac{1}{\eta}}(s)))^2}{f_{\hat{X}}(y_1^{\frac{1}{\eta}}(s))} ds  \right)^{\eta}. \label{y1}
\end{align}

To complete the proof, we note that the expected revenue is given by $R(r_j,t) = r_j^{\frac{1}{\eta}} R(1,t)$ where
\begin{align}
&R(1,t) = \notag \\ &\int_t^T y_1^{\frac{1}{\eta}}(s) \lambda (1 - F_{\hat{X}}(y_1^{\frac{1}{\eta}}(s))) e^{- \int_t^s \lambda (1 - F_{\hat{X}}(y_1^{\frac{1}{\eta}}(s))) dz}ds.
\end{align}
Differentiating the above expression with respect to $t$ gives the following:
\begin{align}
R^{\prime}(1,t) = \lambda (1 - F_{\hat{X}}(y_1^{\frac{1}{\eta}}(t)))(R(1,t) - y_1^{\frac{1}{\eta}}(t)). \label{Revenue_single}
\end{align}
It can be shown that $R(1,t) = \lambda \int_t^T \frac{(1 - F_{\hat{X}}(y_1^{\frac{1}{\eta}}(s)))^2}{f_{\hat{X}}(y_1^{\frac{1}{\eta}}(s))} ds$ satisfies equation~\eqref{Revenue_single}.

\section{Proof of Theorem~} \label{proof_allocation_theorem}
If two VMIs are available, then the revenue can be expressed as follows:
\begin{align}
\int_0^T &\left( P_2(\{r_1,r_2\},t) + R(\{r_1\},t) \right)  h_2(t) dt + \notag \\  &\int_0^T \left( P_1(\{r_1,r_2\},t) + R(\{r_2\},t) \right) h_1(t) dt.
\end{align}
This can be further written as follows:
\begin{align}
&\int_0^T \left(  r_2^{\frac{1}{\eta}} y_2^{\frac{1}{\eta}} + R(r_1,t)\right) \times \notag \\ & \lambda (1 - F_{\hat{X}}(y_2^{\frac{1}{\eta}}(t))) e^{-\int_0^t \lambda (1 - F_{\hat{X}}(y_2^{\frac{1}{\eta}}(s)))ds } dt +  \notag \\ &\int_0^T \left( (r_1^{\frac{1}{\eta}} - r_2^{\frac{1}{\eta}}) y_1^{\frac{1}{\eta}}(t) + R(r_2,t) - R(r_1,t) \right) \times \notag \\
&\lambda (1 - F_{\hat{X}}(y_1^{\frac{1}{\eta}}(t))) e^{ - \int_0^t \lambda (1 - F_{\hat{X}}(y_2^{\frac{1}{\eta}}(s)))ds} dt, \notag \\
&= \int_0^T \left(  r_2^{\frac{1}{\eta}} y_2^{\frac{1}{\eta}} + R(r_1,t)\right) \times \notag \\ &\lambda (1 - F_{\hat{X}}(y_2^{\frac{1}{\eta}}(t))) e^{-\int_0^t \lambda (1 - F_{\hat{X}}(y_2^{\frac{1}{\eta}}(s)))ds } dt + \notag \\ &(r_1^{\frac{1}{\eta}} - r_2^{\frac{1}{\eta}}) \int_0^T \left( y_1^{\frac{1}{\eta}}(t) - R(1,t) \right) \times \notag \\
&\lambda (1 - F_{\hat{X}}(y_1^{\frac{1}{\eta}}(t))) e^{ - \int_0^t \lambda (1 - F_{\hat{X}}(y_2^{\frac{1}{\eta}}(s)))ds} dt.
\end{align}
Let $G(t) = \int_0^t \lambda (1 - F_{\hat{X}}(y_1^{\frac{1}{\eta}}(s))) ds  $ and $H(t) = \int_0^t \lambda (1 - F_{\hat{X}}(y_2^{\frac{1}{\eta}}(s))) ds$. Then, the expression can further be written as follows:
\begin{align}
&\int_0^T  \left(   r_2^{\frac{1}{\eta}} F_{\hat{X}}^{-1} \left( 1 - \frac{H^{\prime}(t)}{\lambda} \right) + r_1^{\frac{1}{\eta}} R(1,t)  \right) H^{\prime}(t) e^{- H(t)} dt  + \notag \\ &(r_1^{\frac{1}{\eta}} - r_2^{\frac{1}{\eta}}) \int_0^T \left(   F_{\hat{X}}^{-1} \left( 1 - \frac{G^{\prime}(t)}{\lambda} \right)  \right) G^{\prime \prime}(t) e^{-H(t)} dt.
\end{align}
Therefore, $L_1(t, H(t), H^{\prime}(t)) = \left(  r_2^{\frac{1}{\eta}} F_{\hat{X}}^{-1} \left( 1 - \frac{H^{\prime}(t)}{\lambda} \right) + r_1^{\frac{1}{\eta}} R(r_1,t)\right) H^{\prime}(t) e^{-H(t) } + (r_1^{\frac{1}{\eta}} - r_2^{\frac{1}{\eta}})\left(  F_{\hat{X}}^{-1} \left( 1 - \frac{G^{\prime}(t)}{\lambda} \right) - R(1,t) \right) G^{\prime}(t) e^{-H(t)}$. 
Computing the Euler-Lagrange equation, $\frac{\partial L_1}{\partial H(t)} - \frac{d}{dt} \frac{\partial L_1}{\partial H^{\prime}(t)} = 0$ results in the following:
\begin{align}
&- (r_1^{\frac{1}{\eta}} - r_2^{\frac{1}{\eta}})  G^{\prime}(t) \left( F_{\hat{X}}^{-1}  \left(1 - \frac{G^{\prime}(t)}{\lambda}\right)  - R(1,t) \right) - \notag \\ &r_1^{\frac{1}{\eta}} R^{\prime}(1,t) -
r_2^{\frac{1}{\eta}}\frac{(H^{\prime}(t))^2  }{ \lambda F_{\hat{X}} \left( F_{\hat{X}}^{-1} \left( 1 - \frac{H^{\prime}(t)}{\lambda} \right) \right)} + \notag\\
&2 r_2^{\frac{1}{\eta}}\frac{ H^{\prime \prime}(t)  }{ \lambda F_{\hat{X}} \left( F_{\hat{X}}^{-1} \left( 1 - \frac{H^{\prime}(t)}{\lambda} \right) \right)}
+ \notag \\
& r_2^{\frac{1}{\eta}} \frac{   F_{\hat{X}}^{\prime} \left( F_{\hat{X}}^{-1} \left( 1 - \frac{H^{\prime}(t)}{\lambda} \right) \right)H^{\prime}(t)  H^{\prime \prime}(t)  }{  \lambda^2 \left( F_{\hat{X}} \left( F_{\hat{X}}^{-1} \left( 1 - \frac{H^{\prime}(t)}{\lambda} \right) \right) \right)^3} = 0,
\end{align}
and $\frac{\partial L_1}{\partial G(t)} - \frac{d}{dt} \frac{\partial L_1}{\partial G^{\prime}(t)} = 0$ results in the following:
\begin{align}
&- H^{\prime}(t) \left( - \frac{G^{\prime} (t)}{ \lambda F_{\hat{X}} \left( F_{\hat{X}}^{-1} \left( 1 - \frac{G^{\prime}(t)}{\lambda} \right) \right) } +  F_{\hat{X}}^{-1} \left( 1 - \frac{G^{\prime}(t)}{\lambda} \right)   - \right. \notag \\ &\left. R(1,t)  \right) - R^{\prime}(1,t) - \frac{ 2 G^{\prime \prime}(t)}{ \lambda F_{\hat{X}} \left( F_{\hat{X}}^{-1} \left( 1 - \frac{G^{\prime}(t)}{\lambda} \right) \right) }     + \notag \\
& 
\frac{f^{\prime}_X \left( F_{\hat{X}}^{-1} \left( 1 - \frac{G^{\prime}(t)}{\lambda} \right) \right) G^{\prime}(t) G^{\prime \prime} (t)}{\lambda^2 f^3_X \left( F_{\hat{X}}^{-1} \left( 1 - \frac{G^{\prime}(t)}{\lambda} \right) \right)  }   = 0.
\end{align}
Eventually, it leads to the following differential equations:
\begin{align}
&-(r_1^{\frac{1}{\eta}} - r_2^{\frac{1}{\eta}}) \lambda \left( 1 - F_{\hat{X}} \left( y_1^{\frac{1}{\eta}}(t) \right) \right)
(y_1^{\frac{1}{\eta}}(t) - R(1,t)) -   \notag \\& r_1^{\frac{1}{\eta}}(t) R^{\prime}(1,t) - r_2^{\frac{1}{\eta}} \frac{\lambda (1 - F_{\hat{X}}(y_2^{\frac{1}{\eta}}(t)))^2 }{ F_{\hat{X}}(y_2^{\frac{1}{\eta}}(t))} - 2 r_2^{\frac{1}{\eta}} \frac{ y_2^{\prime}(t) y_2^{\frac{1}{\eta} - 1}(t)}{ \eta} - \notag \\
&
r_2^{\frac{1}{\eta}}\frac{y_2^{\prime}(t) (1 - F_{\hat{X}}(y_2^{\frac{1}{\eta}}(t)))   F_{\hat{X}}^{\prime}(y_2^{\frac{1}{\eta}}(t)) y_2^{\frac{1}{\eta}-1}(t)   }{ \eta F_{\hat{X}}^2(y_2^{\frac{1}{\eta}}(t))} = 0, \label{first_eq}
\end{align}
and
\begin{align}
& \frac{2 y_1^{\prime}(t) y_1^{\frac{1}{\eta}-1}(t)}{\eta } - R^{\prime}(1,t) - \notag \\ &\frac{ F_{\hat{X}}^{\prime}(y_1^{\frac{1}{\eta}}(t))  (1 - F_{\hat{X}}(y_1^{\frac{1}{\eta}}(t))) y_1^{\prime}(t) y_1^{ \frac{1}{\eta}-1}(t) }{ \eta  f_{\hat{X}}^2(y_1^{\frac{1}{\eta}}(t))   } + \lambda (1-F_{\hat{X}}(y_2^{\frac{1}{\eta}}(t)))  \notag \\ &  \left(  \frac{ (1-F_{\hat{X}}(y_1^{\frac{1}{\eta}}(t)))}{  f_{\hat{X}}(y_1^{\frac{1}{\eta}}(t)) } + y_1^{\frac{1}{\eta}}(t) + R(1,t) \right)  = 0. \label{second_eq}
\end{align}
Now, we show that a solution to these differential equations is given by the solution to the following system of equations:
\begin{align}
y_1(t) = \left( \frac{1 - F_{\hat{X}}(y_1^{\frac{1}{\eta}}(t))}{f_{\hat{X}}(y_1^{\frac{1}{\eta}}(t))} + \lambda \int_t^T \frac{(1 - F_{\hat{X}}(y_1^{\frac{1}{\eta}}(s)))^2}{f_{\hat{X}}(y_1^{\frac{1}{\eta}}(s))} ds \right)^{\eta}, \label{first}
\end{align}
and
\begin{align}
y_2(t) = \Bigg( \frac{1 - F_{\hat{X}}(y_2^{\frac{1}{\eta}}(t))}{f_{\hat{X}}(y_2^{\frac{1}{\eta}}(t))} + &\lambda \int_t^T \frac{(1 - F_{\hat{X}}(y_2^{\frac{1}{\eta}}(s)))^2}{f_{\hat{X}}(y_2^{\frac{1}{\eta}}(s))} ds  -  \notag \\  & R(1,t) \Bigg)^{\eta}.  \label{second}
\end{align}
Differentiating~\eqref{first} with respect to $t$ results in the following:
\begin{align}
2 \frac{y_1^{\frac{1}{\eta}-1} (t) y_1^{\prime}(t) }{\eta} &= - y_1^{\prime}(t) \frac{(1 - F_{\hat{X}}(y_1^{\frac{1}{\eta}}(t))) f^{\prime}(y_1^{\frac{1}{\eta}}(t))}{f_{\hat{X}}^2(y_1^{\frac{1}{\eta}}(t))} - \notag \\ &\frac{ \lambda (1 - F_{\hat{X}}(y_1^{\frac{1}{\eta}}(t)))^2}{f_{\hat{X}}(y_1^{\frac{1}{\eta}}(t))}
\end{align}
Substituting this in~\eqref{second_eq} results in the following:
\begin{align}
\left( \lambda(1 - F_{\hat{X}}(y_1^{\frac{1}{\eta}}(t))) - \lambda (1 - F_{\hat{X}}(y_2^{\frac{1}{\eta}}(t))) \right) \times \notag \\ \left( \frac{1 - F_{\hat{X}}(y_1^{\frac{1}{\eta}}(t))}{  f_{\hat{X}}(y_1^{\frac{1}{\eta}}(t)) } - y_1^{\frac{1}{\eta}}(t) + R(1,t)   \right) = 0,
\end{align}
which is satisfied for all $y_2(t)$ using~\eqref{first}.
Differentiating~\eqref{second} with respect to $t$ results in the following:
\begin{align}
2 \frac{y_2^{\prime}(t) y_2^{\frac{1}{\eta}-1}}{\eta} &= - \frac{(1 - F_{\hat{X}}(y_2(t))F_{\hat{X}}^{\prime}(y_2(t))y_2^{\prime}(t)}{f_{\hat{X}}^2(y_2(t))} -  \notag \\ & \frac{\lambda (1 - F_{\hat{X}}(y_2(t)))^2}{F_{\hat{X}}(y_2(t))} - R^{\prime}(1,t).
\end{align}
Substituting this into~\eqref{second_eq} results in the following:
\begin{align}
-(r_1^{\frac{1}{\eta}} - r_2^{\frac{1}{\eta}}) \lambda (1 - F_{\hat{X}}(y_1^{\frac{1}{\eta}}(t))) (y_1^{\frac{1}{\eta}(t)} - \notag \\ R(1,t)) - R^{\prime}(1,t) \left(r_1^{\frac{1}{\eta}} - r_2^{\frac{1}{\eta}}  \right) = 0
\end{align}
This holds for $y_1(t)$ from~\eqref{Revenue_single}. 

\section{Proof of Theorem~\ref{general_theorem}} \label{proof_general_theorem}
To prove the general case, we show that when two VMIs with unit characteristic are available, the revenue can be computed as follows:
\begin{align}
&R(\{1,1\},t) = \int_t^T \left((y_2^{\frac{1}{\eta}}(t) +  \right. \notag \\ & \left. R(1,s)\right) \lambda (1 - F_{\hat{X}}(y_2^{\frac{1}{\eta}}(s))) e^{- \int_t^s \lambda (1 - F_{\hat{X}}(y_2^{\frac{1}{\eta}}(z))) dz} ds.
\end{align}
Differentiating with respect to $t$ results in 
\begin{align}
&R^{\prime}(\{1,1\},t) = \notag \\ &\lambda (1 - F_{\hat{X}}(y_2^{\frac{1}{\eta}}(t))) (R(\{1,1\}) - y_2^{\frac{1}{\eta}}(t) - R(1,t)),
\end{align}
It can be shown that $R(\{1,1\},t) = \int_t^T \frac{(1 - F_{\hat{X}}(y_2^{\frac{1}{\eta}}(s))^2)}{f_{\hat{X}}(y_2^{\frac{1}{\eta}}(s))} ds$ satisfies the above differential equation using~\eqref{second}.
Using a similar procedure, it can be be shown that in the general case, the optimal threshold solves the following equation:
\begin{align}
y_i(t) = \left( \frac{1 - F_{\hat{X}}(y_i^{\frac{1}{\eta}}(t))}{f_{\hat{X}}(y_i^{\frac{1}{\eta}}(t))} + R(\boldsymbol{1}_i,t) - R(\boldsymbol{1}_{i-1},t)  \right)^{\eta},
\end{align} 
where 
\begin{align}
R(\boldsymbol{1}_i,t) - \lambda \int_t^T \frac{(1 - F_{\hat{X}}(y_i^{\frac{1}{\eta}}(s)) )^2}{f_{\hat{X}}(y_i^{\frac{1}{\eta}}(s))} ds
\end{align}

\ifCLASSOPTIONcaptionsoff
  \newpage
\fi



%


\bibliographystyle{IEEEtran}
\bibliography{references}

%


\begin{IEEEbiography}
	[{\includegraphics[width=1in,height=1.25in,clip,keepaspectratio]{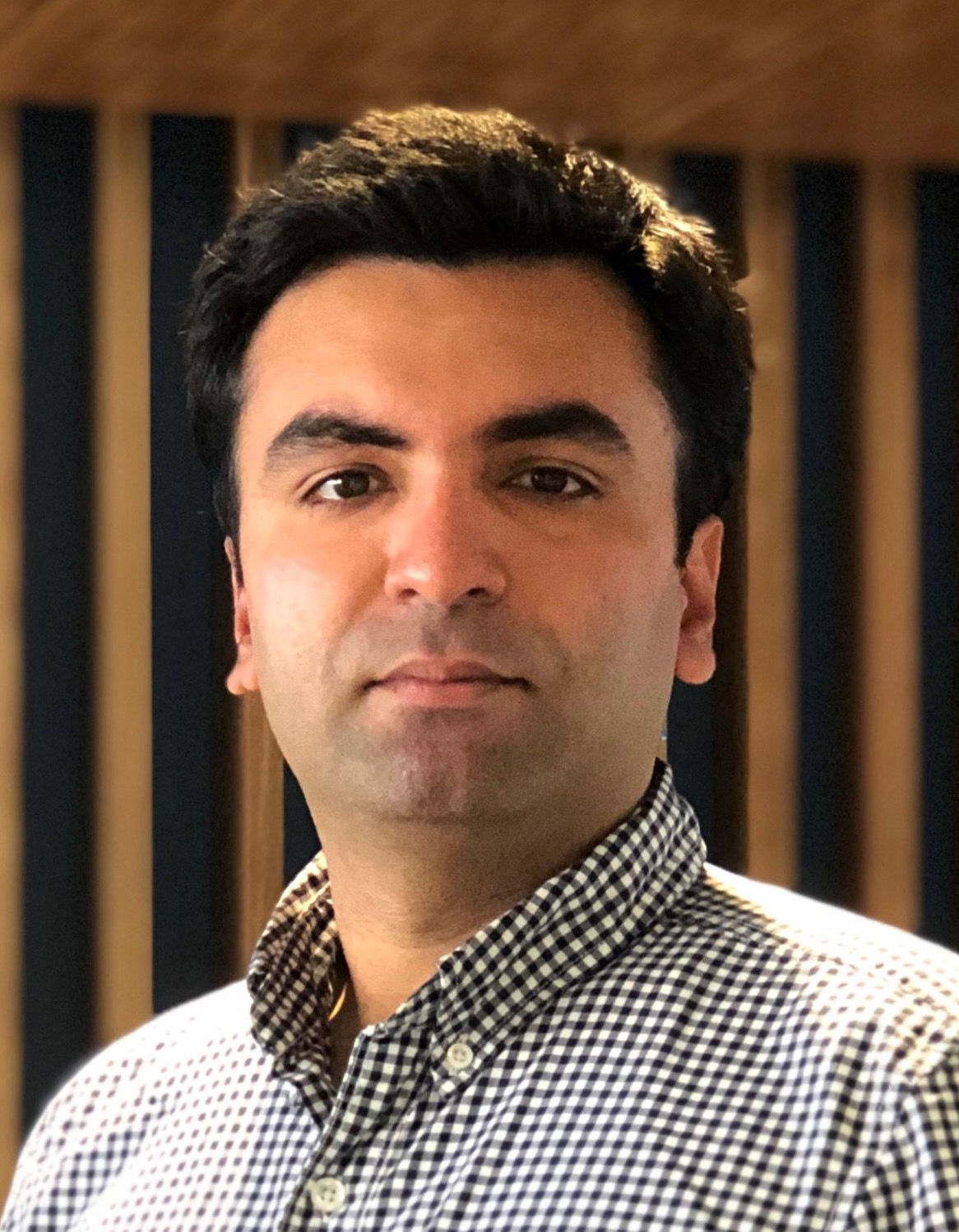}}]{Muhammad Junaid Farooq} (S'15) received the B.S. degree in electrical engineering from the School of Electrical Engineering and Computer Science (SEECS), National University of Sciences and Technology (NUST), Islamabad, Pakistan in 2013, the M.S. degree in electrical engineering from the King Abdullah University of Science and Technology (KAUST), Thuwal, Saudi Arabia in 2015, and the Ph.D. degree in electrical engineering from the Tandon School of Engineering, New York University, Brooklyn, NY in 2020. From 2015 to 2016, he was a Research Assistant with the Qatar Mobility Innovations Center (QMIC), Qatar Science and Technology Park (QSTP), Doha, Qatar. Currently, he is an Assistant Professor with the Department of Electrical and Computer Engineering, College of Engineering and Computer Science, University of Michigan-Dearborn. His research interests include modeling, analysis and optimization of wireless communication systems, cyber-physical systems, and the Internet of things. He is a recipient of the President's Gold Medal for academic excellence from NUST, the Ernst Weber Fellowship Award for graduate studies, the Athanasios Papoulis Award for teaching excellence, and the Dante Youla Award for research excellence from the Department of Electrical \& Computer Engineering (ECE) at NYU Tandon School of Engineering.\end{IEEEbiography}


\begin{IEEEbiography}[{\includegraphics[width=1in,height=1.25in,clip,keepaspectratio]{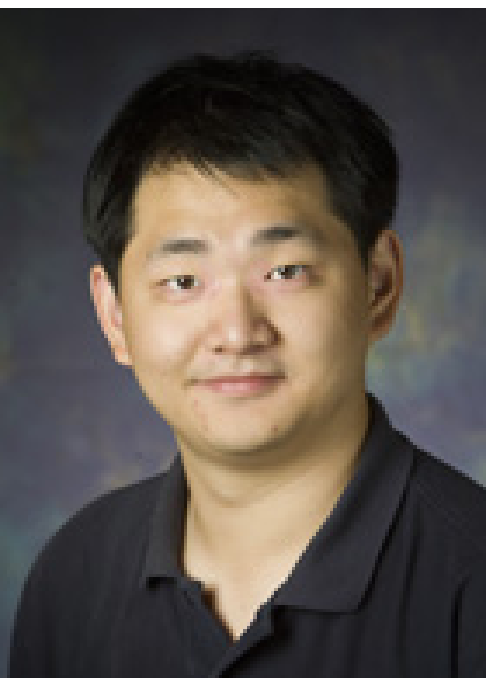}}]{Quanyan Zhu} (S'04, M'12) (SM’02-M’14) received B. Eng. in Honors Electrical Engineering from McGill University in 2006, M. A. Sc. from the University of Toronto in 2008, and Ph.D. from the University of Illinois at Urbana-Champaign (UIUC) in 2013. He is currently an associate professor at the Department of Electrical and Computer Engineering, New York University (NYU). He is an affiliated faculty member of the Center for Cyber Security (CCS) and the Center for Urban Science and Progress (CUSP) at NYU. His current research interests include game theory, machine learning, cybersecurity and deception, network optimization and control, Internet of Things, and cyber-physical systems.
\end{IEEEbiography}




\end{document}